\documentclass[aps,prd,preprint,preprintnumbers,showpacs,showkeys,nofootinbib,superscriptaddress]{revtex4-1}

\usepackage{amsmath}
\usepackage{amssymb}
\usepackage{dcolumn}	
\usepackage{bm}			
\usepackage{bbm} 		
\usepackage{multirow}
\usepackage{blkarray}
\usepackage{booktabs}
\usepackage{feynmp}
\usepackage[usenames,dvipsnames]{xcolor}
\definecolor{hgreen}{rgb}{0,.3,0}
\definecolor{hred}{rgb}{.3,0,0}
\definecolor{hblue}{rgb}{0,0,.3}
\definecolor{LightGray}{gray}{0.95}
\usepackage[colorlinks=true,linkcolor=hblue,citecolor=hgreen,filecolor=hblue,
urlcolor=hred]{hyperref}
\makeatletter
\def\endfmffile{%
	\fmfcmd{\p@rcent\space the end.^^J%
		end.^^J%
		endinput;}%
	\if@fmfio
	\immediate\closeout\@outfmf
	\fi
	\ifnum\pdfshellescape=\@ne
	\immediate\write18{mpost \thefmffile}%
	\fi}
\makeatother

\usepackage{hyperref,graphicx}           

\usepackage{pbox}
\usepackage{caption}
\usepackage{subcaption}
\usepackage{xcolor}
\usepackage[utf8]{inputenc}

\definecolor{Red}{rgb}{1.,0.,0.}
\definecolor{Grn}{rgb}{0.,0.75,0.}
\definecolor{Blu}{rgb}{0.,0.,1.}

\newcommand{\lag}{\mathcal{L}}

\graphicspath{{./figs/}}

\begin{document}

\preprint{DESY 18-187}

\title{Singleton Portals to the Twin Sector}

\author{Fady Bishara}
\affiliation{Deutsches Elektronen-Synchrotron (DESY), D-22607 Hamburg, Germany}
\affiliation{Rudolf Peierls Centre for Theoretical Physics, University of Oxford OX1 3NP Oxford, United Kingdom}
\author{Christopher B. Verhaaren}
\affiliation{Center for Quantum Mathematics and Physics (QMAP), Department of Physics,\\ University of California, Davis, CA, 95616-5270 USA}

\date{\today}
\begin{abstract}
The mirror twin Higgs framework allows for a natural Higgs mass while being consistent with collider bounds on colored symmetry partners to standard model quarks. This mechanism relies crucially on a discrete symmetry which relates each standard model field to a mirror partner. These partners are charged under gauge groups identical to, but distinct from, those in the standard model. The minimal twin Higgs scenario provides only one low-energy connection between the visible and twin sectors, the light Higgs boson. We present a new class of portals connecting the two sectors, using fields that have \emph{no twin} partner under the discrete symmetry. Scalar, fermion, and vector states may provide such \emph{singleton} portals, each with unique features and experimental signatures. The vector portal, in particular, provides a variety of renormalizable interactions relevant for the LHC. We provide concrete constructions of these portals and determine their phenomenology and opportunities to probe the twin sector at the LHC. We also sketch a scenario in which the structure of the twin sector itself can be tested. 
\end{abstract}



\maketitle

\section{Introduction\label{sec.intro}}
While the standard model (SM) is the dominant paradigm in particle physics it is not the complete theory of nature. Its limitations (ignoring gravity) fall into three classes. The first concerns its structure, such as the number of fermion generations, the hierarchical pattern of quark mixing, the anarchic pattern of lepton mixing, and the apparent minimality of electroweak symmetry breaking (EWSB) and fermion mass generation. The second class pertains to its failure to explain experimental facts such as neutrino masses and mixing, dark matter, and the observed baryon asymmetry of the Universe. The final class relates to apparently finely chosen parameters, like the smallness of the effective QCD $\theta$ angle and the hierarchy between the electroweak and Planck scales. 

Extending of the SM to account for these shortcomings is a driver of theoretical particle physics research. While the motivations can appear theoretical, success is claimed only when theory and experiment agree. To this end, the Large Hadron Collider (LHC) continues to search for new particles and interactions that might explain the outstanding puzzles in particle physics.
Following the discovery of the SM-like Higgs boson~\cite{Aad:2012tfa,Chatrchyan:2012xdj}, one of the top priorities of the LHC is to examine its properties. This includes measuring Higgs couplings to SM fields, as well as using the Higgs itself to probe beyond the SM (BSM).  

The hierarchy between the electroweak scale, which sets the Higgs mass, and the Planck scale is arguably a strong link between Higgs and BSM physics. The LHC has completed many powerful and sophisticated searches for new particles that might begin to explain this hierarchy through Higgs compositeness~\cite{Kaplan:1983fs,Kaplan:1983sm,Georgi:1984af,Dugan:1984hq} or a new symmetry~\cite{Fayet:1977yc,Dimopoulos:1981zb,ArkaniHamed:2001nc}. So far, however, no hints have appeared. 

This further motivates the neutral naturalness framework, of which the twin Higgs~\cite{Chacko:2005pe} is the prime example, where the Higgs mass is naturally light while being consistent these null results. This class of symmetry based solutions to the hierarchy problem employ a discrete symmetry to ensure that the symmetry partners of SM fields do not carry SM color. The partner fields may~\cite{Burdman:2006tz,Cai:2008au,Cohen:2015gaa,Gherghetta:2016bcc,Xu:2018ofw} or may not~\cite{Chacko:2005pe,Barbieri:2005ri,Poland:2008ev,Craig:2014aea,Craig:2014roa,Craig:2015xla,Batell:2015aha,Serra:2017poj,Csaki:2017spo,Cohen:2018mgv,Cheng:2018gvu} carry SM electroweak charges, which determines much of their collider phenomenology~\cite{Barbieri:2005ri,Burdman:2008ek,Burdman:2014zta,Curtin:2015fna,Curtin:2015bka,Burdman:2015oej,Chacko:2015fbc,Thrasher:2017rpa,Lichtenstein:2018kno}. 

The minimal mirror twin Higgs (MTH) framework has been modified and expanded in various ways~\cite{Craig:2015pha,Beauchesne:2015lva,Bai:2015ztj,Craig:2016kue,Yu:2016cdr,Yu:2016swa,Barbieri:2016zxn,Yu:2016cdr,Badziak:2017syq,Badziak:2017kjk,Badziak:2017wxn}. Significant work has gone into reconciling the twin sector with cosmological data~\cite{Schwaller:2015tja,Kilic:2015joa,Barbieri:2016zxn,Craig:2016lyx,Chacko:2016hvu,Csaki:2017spo}. Indeed, it has been demonstrated that this framework can address many cosmological phenomena~\cite{Freytsis:2016dgf,Chacko:2018vss,Fujikura:2018duw}, such as the nature of dark matter~\cite{Craig:2015xla,Garcia:2015loa,Garcia:2015toa,Farina:2015uea,Hochberg:2018vdo,Cheng:2018vaj}, structure anomalies~\cite{Prilepina:2016rlq}, and baryogenesis~\cite{Farina:2016ndq}. Connections to flavor~\cite{Csaki:2015gfd,Barbieri:2017opf} and the neutrino sector~\cite{Batell:2015aha,Bai:2015ztj,Chacko:2016hvu,Csaki:2017spo} have also been explored. While this work focuses on the low energy aspects of the twin Higgs, assuming an EFT cutoff of a few TeV, both supersymmetric~\cite{Falkowski:2006qq,Chang:2006ra,Craig:2013fga,Katz:2016wtw,Asadi:2018abu} and composite~\cite{Batra:2008jy,Geller:2014kta,Barbieri:2015lqa,Low:2015nqa} UV completions are possible and can lead to interesting signals at future colliders~\cite{Cheng:2015buv,Cheng:2016uqk,Contino:2017moj}.

The MTH model has only one guaranteed low-energy portal between the two sectors, the Higgs itself. Kinetic mixing between SM and twin hypercharge can provide another portal and has been explored in the context of dark matter signals~\cite{Craig:2015xla,Garcia:2015toa}. The heavy twin Higgs can be another portal when the UV completion is weakly coupled, and its collider signals have also been explored~\cite{Ahmed:2017psb,Chacko:2017xpd}.

These particles can connect the two sectors because they carry no gauge charges. In this case the SM field $\psi_A$ can mix with its twin partner $\psi_B$ without violating any symmetries 
\begin{equation}
    \psi_A\psi_B\xrightarrow{\mathbbm{{Z}_2}}\psi_B\psi_A.
\end{equation}
This mixing allows the physical eigenstates to be a linear combination of the SM and twin fields thereby linking the sectors. 

However, there is another class of BSM states that can connect the two sectors. These particles have no partner under the discrete symmetry, but change by at most a phase. The $\mathbbm{{Z}_2}$ symmetry of the twin Higgs model requires the phase to be either 0 or $\pi$ 
\begin{equation}
    \psi\xrightarrow{\mathbbm{{Z}_2}}\pm\psi.
\end{equation}
Because the particles have no twin partner under the $\mathbb{Z}_2$ we refer to them as \emph{singletons}. Schematically, their coupling is
\begin{equation}
    \psi\left(\mathcal{O}_A\pm \mathcal{O}_B \right),\label{e.Opers}
\end{equation}
where $\mathcal{O}_{A,B}$ is a set of operators in the $A$ or $B$ sector.
Since the $\mathbb{Z}_2$ symmetry exchanges the operators, $\mathcal{O}_A\leftrightarrow\mathcal{O}_B$, the relative sign in Eq.~\eqref{e.Opers} is fixed by the phase of $\psi$ such that the interaction is invariant under $\mathbb{Z}_2$.

Since the $A$ and $B$ fields transform under distinct gauge groups, the singletons must be SM and twin gauge singlets. This limits the possible renormalizable interactions of scalar and fermionic singletons to SM fields. However, a new $U(1)_X$ gauge symmetry under which both the SM and twin sectors are charged is much less constrained.

In this work we examine these singleton portals, with particular emphasis on LHC signals. We begin, in Sec.~\ref{s.singleport}, by examining the allowed forms of scalar, fermion, and vector portals, focusing on renormalizable interaction. Then, in Sec.~\ref{sec.model} we study the LHC signals of several benchmark vector singletons and constraints from other experiments. The details of the model and some formulae useful to the analysis are given in Apps.~\ref{a.kinmix}~and~\ref{app:useful}. In Sec.~\ref{sec:conclusion} we sketch some directions of further investigation and conclude.

\section{Singleton Portals\label{s.singleport}}
\subsection{Scalar portal}
The simplest type of singleton portal is a scalar, $\phi$. The only gauge invariant renormalizeable couplings of $\phi$ to SM and twin fields that preserved the discrete symmetry are
\begin{align}
 \Delta\lag=&\kappa\phi\left(  \left|H_{A}\right|^2\pm\left|H_B \right|^2 \right)+ \lambda_{H\phi}\left|\phi\right|^2\left(  \left|H_{A}\right|^2+\left|H_B \right|^2 \right),
 \label{eq:scalar-portal}
\end{align}
where the minus sign applies when $\phi$ is odd under the $\mathbb{Z}_2$. 
 If $\phi$ is odd, and acquires a vacuum expectation value (VEV), the first term in Eq.~\eqref{eq:scalar-portal} spontaneously breaks the $\mathbb{Z}_2$ symmetry.\footnote{See~\cite{Beauchesne:2015lva} for a model of spontaneous $\mathbb{Z}_2$ breaking without a singleton field.} This provides an attractive origin of the soft $\mathbb{Z}_2$ breaking required by Higgs coupling measurements~\cite{Burdman:2014zta}. 

If $\phi$ transforms nontrivially under any symmetry other than the twin $\mathbb{Z}_2$, then only the second operator in Eq.~\eqref{eq:scalar-portal} is allowed. This includes a simple $\phi\to-\phi$ symmetry that might stabilize $\phi$ against decay, making it a possible dark matter candidate.\footnote{Being odd under the twin $\mathbb{Z}_2$ does not stabilize $\phi$, as the first term in Eq.~\eqref{eq:scalar-portal} makes clear.} However, this $|\phi|^2$ interaction is also $SU(4)$ symmetric in the Higgs fields, and consequently does not contribute to the potential of pseudo-Nambu-Goldstone bosons (pNGBs) like the physical Higgs. Therefore, in this case $\phi$ and the physical Higgs have no renormalizable interactions. Similarly, when $\phi$ is even under the twin $\mathbb{Z}_2$ the first term in Eq.~\eqref{eq:scalar-portal} respects the global symmetry of the Higgs potential, providing no interactions between the physical Higgs and $\phi$. 

If $\phi$ is odd under the twin $\mathbb{Z}_2$, and is not protected by some symmetry from large quantum corrections, we expect its mass to be near the cutoff of a few TeV. To leading order a heavy $\phi$ merely modifies the terms in the Higgs potential and we can simply integrate it out of the low energy theory. Note, however, that near the cutoff can mean a loop factor below. Because Eq.~\eqref{eq:scalar-portal} includes all the interactions with low energy fields, Higgs loops and self interactions dominate corrections to the $\phi$ mass. For an order one $\lambda_{H\phi}$ and a 5 TeV cutoff the mass of $\phi$ can naturally be a few hundred GeV.

Clearly, the most interesting case for LHC signals is for $\phi$ to be odd under the twin $\mathbb{Z}_2$ and not too heavy. Then its VEV spontaneously breaks the symmetry, providing the necessary soft $\mathbb{Z}_2$ breaking mass term in the Higgs potential. To see this explicitly, we write the $\mathbb{Z}_2$ preserving Higgs potential, without including $\phi$ and using the notation of~\cite{Barbieri:2005ri}, as
\begin{equation}
    V_H=-\mu^2\left(\left|H_{A}\right|^2+\left|H_{B}\right|^2\right)+\lambda_H\left(\left|H_{A}\right|^2+\left|H_{B}\right|^2 \right)^2+\delta\left(\left|H_{A}\right|^4+\left|H_{B}\right|^4 \right),
\end{equation}
where the potential has an approximate $SU(4)$ symmetry and leads to nearly equal VEVs for $H_A$ and $H_B$ if we take  $0<\delta\ll 1$.

We use a nonlinear parameterization of the Higgs field, see~\cite{Burdman:2014zta}, neglecting the heavy radial mode. We work in unitary gauge where are all but one the pNGBs of the broken $SU(4)$ have been eaten, and find
\begin{align}
    \left|H_{A}\right|^2=f^2\sin^2\left(\frac{v+h}{f\sqrt{2}}\right), \ \ \ \
    \left|H_{B}\right|^2=f^2\cos^2\left(\frac{v+h}{f\sqrt{2}}\right).
\end{align}
 Here $f$ is the $SU(4)$ breaking VEV, $v$ parameterizes how much of the VEV is in either sector, and $h$ is the physical Higgs boson. By defining $\vartheta\equiv v/(f\sqrt{2})$ we find the VEVs in each sector are
\begin{equation}
    v_A=f\sqrt{2}\sin\vartheta, \ \ \ \ v_B=f\sqrt{2}\cos\vartheta, \label{e.HiggsVeVs}
\end{equation}
where $v_A=v_\text{EW}=246$ GeV. In this language the Higgs potential is
\begin{equation}
    V_H=-\mu^2f^2+\lambda_H f^4+ f^4\frac{\delta}{4}\left[3+\cos\frac{2\sqrt{2}(v+h)}{f} \right].
\end{equation}
By requiring that there be no tadpole term we find $\sin4\vartheta=0$, which since must $\vartheta\neq0$ to have nonzero $v_A$ implies $\vartheta=\pi/4$, and hence $v_A=v_B$. We also find the Higgs mass is
\begin{equation}
    m_h^2=-2f^2\delta\cos4\vartheta=2\delta f^2,
\end{equation}
which, as expected, is controlled by $\delta$, the global symmetry breaking parameter of the potential.

In the twin Higgs set-up Higgs couplings $g_h$ satisfy
\begin{equation}
    g_h=g_{h\text{SM}}\cos\vartheta,
\end{equation}
where $g_{h\text{SM}}$ is the SM coupling~\cite{Burdman:2014zta}. Therefore, having equal VEV in each sector implies the Higgs' couplings to visible fields are half the SM prediction, which is ruled out by LHC measurements. This result is modified by introducing a $\mathbb{Z}_2$-odd singleton,
\begin{equation}
    \phi=v_\phi+\varphi,
\end{equation}
with VEV $v_\phi$ and potential
\begin{equation}
    V_\phi=-m^2\phi^2+\lambda_\phi\phi^4.
\end{equation}
The interaction between the scalars in Eq.~\eqref{eq:scalar-portal} become
\begin{equation}
 \kappa\left(v_\phi+\varphi\right) f^2\left[\cos2\vartheta-\frac{\sqrt{2}\sin2\vartheta}{f}h-\frac{\cos2\vartheta}{f^2}h^2+\ldots \right]+\lambda_\phi f^2\left(v_\phi+\varphi\right)^2.
\end{equation}
This modifies the Higgs' no-tadpole constraint to
\begin{equation}
    \cos2\vartheta=-\frac{\kappa v_\phi}{\delta f^2}.
\end{equation}
Experiment requires $\vartheta\lesssim 1/3$, so the above relation requires $\kappa<0$. The Higgs mass becomes
\begin{equation}
    m_h^2=2\delta f^2\left(1-\frac{v_\phi^2\kappa^2}{\delta^2f^4} \right)=2\delta f^2\sin^22\vartheta=4\delta v_A^2\cos^2\vartheta.
\end{equation}

\begin{figure}[t]
	\includegraphics[width=0.45\textwidth]{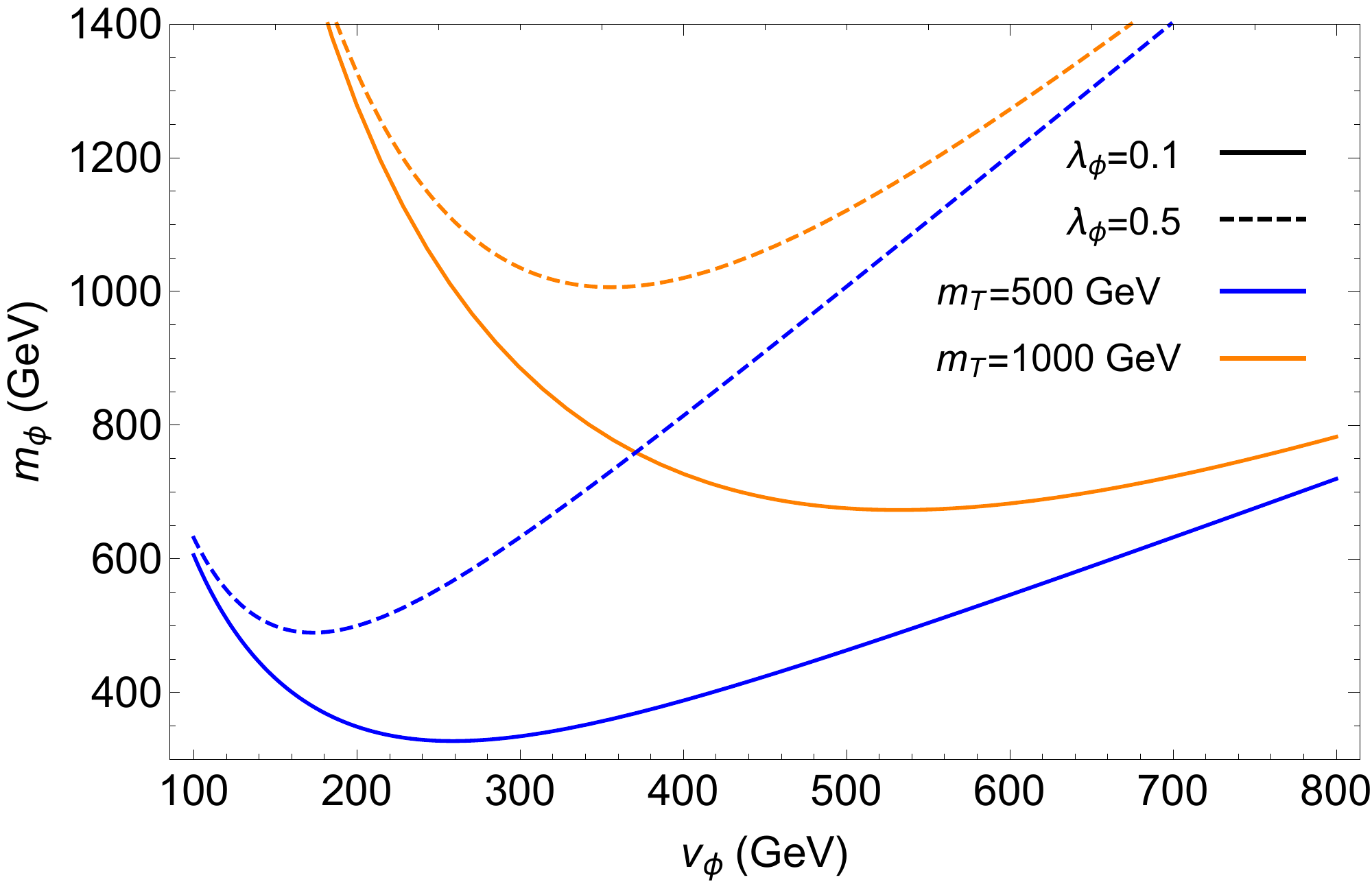}
	\includegraphics[width=0.45\textwidth]{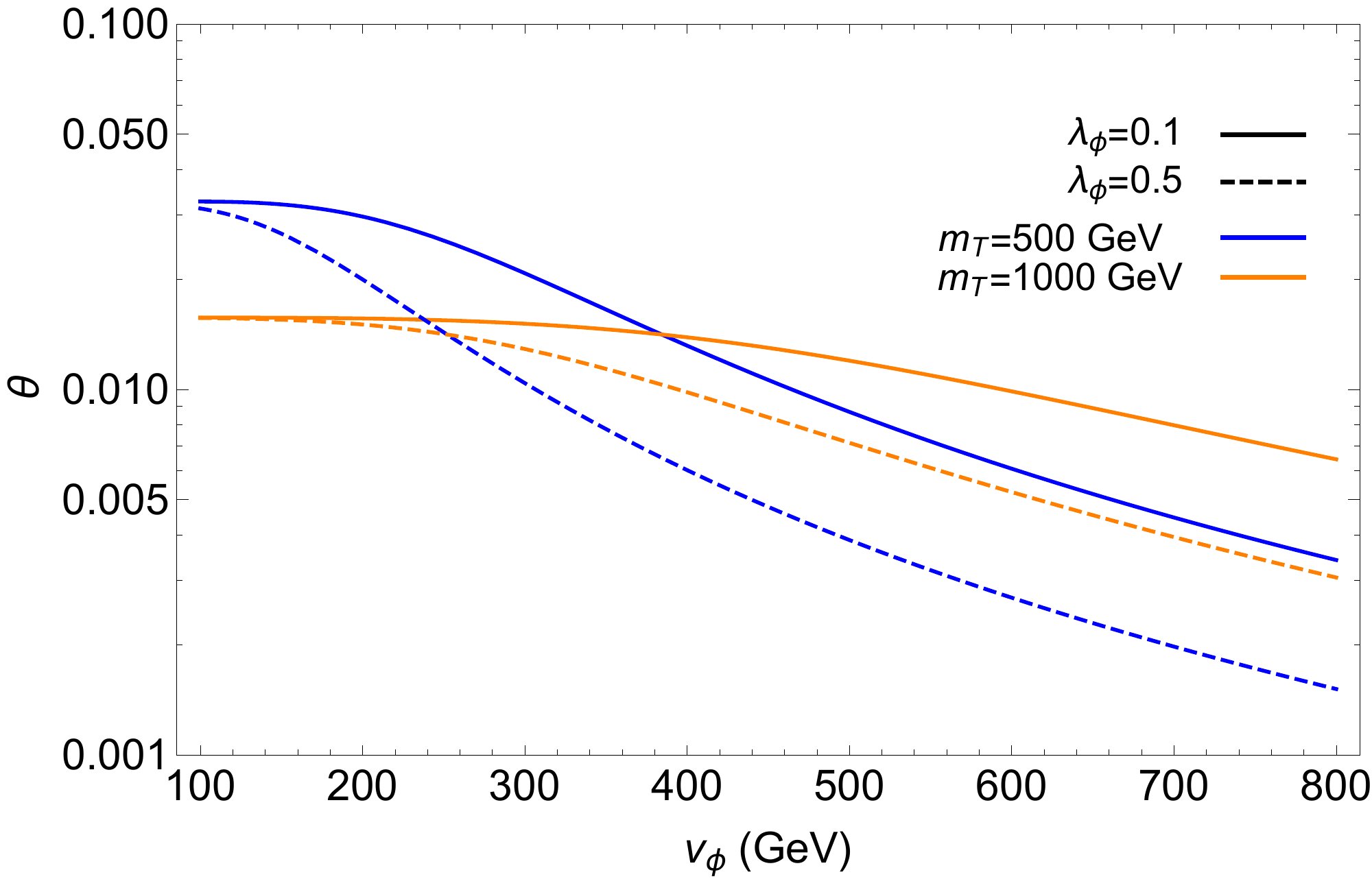}
	\caption{On the left (right) we plot the mass (mixing angle with the Higgs) of the singleton scalar for twin top mass of 500 (1000) GeV in blue (orange). The quartic coupling $\lambda_\phi$ is 0.1 (0.5) for the solid (dashed) lines. The mass is typically at least a few hundred GeV and can be much heavier. The the mixing angle is always small. }
	\label{f.ScalarMass}
\end{figure}

The constraint that $\varphi$ have no tadpole and the mass of $\varphi$ are
\begin{align}
    0=&4\lambda_\phi v_\phi^2-2m^2+2\lambda_{H\phi}f^2-\frac{\kappa^2}{\delta},\\
    m_\varphi^2=&8v_\phi^2\lambda_\phi+\frac{\kappa^2}{\delta}=8v_\phi^2\lambda_\phi+\frac{m_h^2v_A^2}{4v_\phi^2}\frac{\cot^22\vartheta}{\sin^2\vartheta}.
\end{align}
There is also a mass mixing term between the scalars
\begin{equation}
    -h\varphi\sqrt{2}\kappa f\sin2\vartheta=h\varphi \frac{m_h^2v_A\cot2\vartheta}{2v_\phi \sin\vartheta}\,.
\end{equation}
This is diagonalized by a simple matrix
\begin{equation}
    \left(\begin{array}{c}
    \widehat{h}\\
    \widehat{\varphi}
    \end{array}\right)=\left(\begin{array}{cc}
    \cos\theta & -\sin\theta\\
    \sin\theta & \cos\theta
    \end{array}\right)\left(\begin{array}{c}
    h\\
    \varphi
    \end{array}\right),
\end{equation}
where
\begin{equation}
    \tan2\theta=\frac{m_h^2v_A\cot2\vartheta}{2v_\phi \sin\vartheta(m_\varphi^2-m_h^2)},
\end{equation}
and the mass eigenvalues are
\begin{equation}
    m^2_{\widehat{\varphi},\widehat{h}}=\frac12\left(m_\varphi^2+m_h^2\pm\sqrt{\left(m_\varphi^2-m_h^2\right)^2+\frac{m_h^4v_A^2\cot^22\vartheta}{4v_\phi^2\sin^2\vartheta}} \right)\,.
\end{equation}
From Fig.~\ref{f.ScalarMass} we find that our estimate of $\varphi$ having a few hundred GeV mass is borne out for larger $\vartheta$, but as this angle decreases, $m_\varphi$ increases. The right plot makes clear that the mixing angle $\theta$ is always quite small.

Including the effects of $\phi$ we find the couplings of $\widehat{h}$ and $\widehat{\varphi}$ to SM fields
\begin{equation}
    g_{\widehat{h}}=g_{h\text{SM}}\cos\vartheta\cos\theta, \ \ \ \ g_{\widehat{\varphi}}=-g_{h\text{SM}}\cos\vartheta\sin\theta.
\end{equation}
This means that production of the $\widehat{\varphi}$ proceeds exactly as the Higgs, but with the production rate reduced by $\sin^2\theta\lesssim 10^{-4}$. Unfortunately, this reduction in rate makes $\widehat{\varphi}$ very challenging to discover, even at next generation colliders. The most promising channel at a hadron collider would likely be production through gluon fusion and decay to $ZZ$. However, even at 100 TeV the projected bounds for such a scalar~\cite{Buttazzo:2015bka} remain several orders of magnitude above the $\widehat{\varphi}$ cross section. A high energy lepton collider can exploit the $\widehat{\varphi}\to\widehat{h}\widehat{h}\to bbbb$ final state~\cite{Chacko:2017xpd,Buttazzo:2018qqp}, but here too, the cross section is prohibitively small. Therefore, while the neutral singleton provides a simple origin of the soft $\mathbb{Z}_2$ breaking in the Higgs potential, its experimental signatures may be too faint to see in the near future.

\subsection{Fermionic portal}
The most obvious fermionic singleton is the gauge singlet right-handed neutrino $\nu_R$. This can couple as 
\begin{equation}
-\Delta\mathcal{L}=
\left(\overline L_A\,Y_A\,\nu_R\right)H_A\pm 
\left(\overline L_B\,Y_B\,\nu_R\right)H_B +
\frac{m_R}{2}\overline{\nu}_R^c\nu_R+ \text{H.c.} ~,
\end{equation}
where the minus sign applies if $\nu_R$ is odd under the $\mathbbm{Z}_2$. The Yukawa couplings $Y_{A,B}$ are complex rectangular matrices in flavor space, which are forced to be identical by the discrete symmetry, so we simply write them as $Y_\nu$. The $\nu_R$'s large Majorana mass term, $m_R$, leads to a seesaw~\cite{Minkowski:1977sc,Yanagida:1979as,GellMann:1980vs,Glashow1980,Mohapatra:1979ia} in both sectors.

However, $\nu_R$ also induces mixing between the SM and twin neutrinos, creating the mass matrix
\begin{equation}
\big(\nu_A,\;\nu_B\big)\,\frac{Y_\nu^2}{m_R}\left( \begin{array}{cc}
v_A^2 & \pm v_Av_B\\
\pm v_Av_B & v_B^2
\end{array}\right)\left( \begin{array}{c}
\nu_A\\
\nu_B
\end{array}\right),
\end{equation}
where we see the magnitudes of $Y_\nu$ and $m_R$ have factored out. The physical eigenstates are
\begin{align}
    \nu_+=&\nu_B\cos\theta_\nu\pm\nu_A\sin\theta_\nu,\\
    \nu_-=&\nu_A\cos\theta_\nu\mp\nu_B\sin\theta_\nu,
\end{align}
where the mixing angle is defined by
\begin{equation}
\sin\theta_\nu=\frac{v_A}{\sqrt{v_A^2+v_B^2}}\,.
\end{equation}
The mass eigenvalues are
\begin{equation}
   m_{-}=0,\ \ \ \ m_+=\frac{Y_\nu^2}{M_R}\left(v_A^2+v_B^2\right),
\end{equation}
where $v_i$ is the VEV of $H_i$ given in Eq.~\eqref{e.HiggsVeVs}. 

No matter the values of $Y_\nu$ and $m_R$, the mixing angle is completely determined by the hierarchy of VEVs in the two sectors. Collider bounds on Higgs couplings imply that $v_B\gtrsim 3v_A$, while requiring the tuning be no worse than 10\% implies $v_B\lesssim 6v_A$~\cite{Burdman:2014zta}. Therefore, the interesting range of mixing is
\begin{equation}
    \sin^2\theta_\nu\in[0.03,0.1].\label{e.twinNuMix} 
\end{equation}

As pointed out in~\cite{Csaki:2017spo} this mixing between the $A$ and $B$ neutrinos can be used to reduce the contribution to $\Delta N_\text{eff}$ from the twin sector, which CMB measurements constrain to $\Delta N_\text{eff}\lesssim0.3$ at 95\% confidence~\cite{Aghanim:2018eyx}. However, the bounds on neutrino mixing can be quite constraining. In particular, meson decays bound neutrino mixing in these cases, see for instance~\cite{Batell:2017cmf}, often an order of magnitude beyond the prediction given in Eq.~\eqref{e.twinNuMix}. The only exception is if the singleton only connects the $\tau$ neutrinos. The limits here are weaker, and this level of mixing can be accommodated if the twin neutrinos are not heavier than a few hundred MeV. This is easily accomplished by $M_R\sim$ TeV and $Y_\nu$ order one.

One might hope to avoid this large mixing by using the inverse seesaw mechanism~\cite{Mohapatra:1986aw,Mohapatra:1986bd}. In the usual inverse seesaw set-up, a Dirac partner $\overline{N}$ is given a TeV scale mass $M$ with the right-handed neutrino, and a small Majorana mass $\mu$,
\begin{equation}
    \overline{L}Y_\nu\nu_R H+M\overline{N}\nu_R +\frac{\mu}{2}\overline{N}\,\overline{N}^c+\text{H.c.}\,.
\end{equation}
The mass of the left-handed neutrino is then
\begin{equation}
    m_\nu\sim\frac{Y_\nu^2v_{\text{EW}}^2}{2M^2}\frac{\mu}{2},
\end{equation}
where $v_\text{EW}=246$ GeV is the Higgs VEV. This gives a neutrino mass in the tens of meV range for $M\sim$ TeV, $\mu\sim$ keV, and $Y\sim10^{-2}$. The heavy neutrinos have $M$ scale masses.

Because both $\nu_R$ and $\overline{N}$ are gauge singlets, either can be singletons under the twin $\mathbb{Z}_2$. However, as we saw in the standard seesaw, taking $\nu_R$ as a singleton leads to large mixing between the SM and twin neutrinos. This remains true in the inverse seesaw, but now the SM neutrino states are exactly masses and the twin neutrinos have mass
\begin{equation}
    m_{\nu B}\sim\frac{Y_\nu^2(v_A^2+v_B^2)}{2M^2}\frac{\mu}{2}.
\end{equation}

By choosing $Y_\nu$ and $\mu$ appropriately, since $(v_A^2+v_B^2)\lesssim$ 1 TeV both $Y_\nu$ and $\mu$ need to be smaller than the standard case, the twin neutrinos can be made light enough to escape the bounds from meson decays. These neutrinos are light enough to lead to tension with $\Delta N_\text{eff}$ bounds, but this can be overcome in other ways such as asymmetric reheating of the two sectors~\cite{Chacko:2016hvu}. In this regime future precision $\beta$-decay experiments will probe down to eV masses through combination of endpoint~\cite{Mertens:2014nha} and kink~\cite{Formaggio:2011jg} measurements of the $\beta$-decay spectrum. The values of the mixing given in Eq.~\eqref{e.twinNuMix}, at least to the electron neutrino, are projected to be probed by these experiments. 

Finally, we can also choose the right-handed neutrinos to be twin partners $\nu_{RA}$ and $\nu_{RB}$ which have Dirac masses with the singleton $\overline{N}$. The Lagrangian is
\begin{equation}
-\Delta\mathcal{L}=
\left(\overline L_A\,Y_A\,\nu_{RA}\right)H_A+ 
\left(\overline L_B\,Y_B\,\nu_{RB}\right)H_B +
M\overline{N}\left(\nu_{RA}\pm\nu_{RB} \right) +\frac{\mu}{2}\overline{N}\,\overline{N}^c +\text{H.c.} ~,
\end{equation}
with the sign of the Dirac mass term determined by the twin parity of $\overline{N}$. The lightest state is mostly SM neutrino, and has a mass
\begin{equation}
    m_\nu\sim\frac{Y^2v_A^2}{1+v_A^2/v_B^2 }\frac{\mu}{4M^2}.
\end{equation}
Though the mixing of this neutrino with the $M$ mass states is small, there is large mixing with the two states with mass
\begin{equation}
    m_{-}\sim\frac{\sqrt{v_A^2+v_B^2}}{2},
\end{equation}
which is at or just below the TeV scale. This again leads to the mixing in Eq.~\eqref{e.twinNuMix} and because the mixing states are heavy, leads to the same constraints as in the standard seesaw. 

In summary, constraints on neutrino mixing limit the possible fermionic singleton portals. However, there are few cases, like $\tau$ flavored singletons or inverse seesaw with singleton right-handed neutrino that remain viable. The latter scenario offers signals that will be probed soon by the measurement of the $\beta$-decay spectrum, at least in the electron flavored case.

\subsection{Vector portal}
A singleton vector $X_\mu$ can link the SM to the twins through kinetic mixing with the $U(1)_Y$ gauge fields of each sector
\begin{align}
    -\frac{\varepsilon}{2}\left(B_A^{\mu\nu}\pm B_B^{\mu\nu}\right)X_{\mu\nu},\label{e.kinMix}
\end{align}
where $B_{\mu\nu}$ is the field strength of the hypercharge gauge boson $B_\mu$ and similarly for $X_\mu$. The sectors can also be connected by direct coupling of $X_\mu$ to matter currents 
\begin{equation}
    g_X X^\mu\left[\overline{f}_A\gamma_\mu\left(C_V+\gamma_5 C_A \right)f_A\pm \overline{f}_B\gamma_\mu\left(C_V+\gamma_5 C_A \right) f_B\right].\label{e.Vcoup}
\end{equation}
In each case the sign between the two terms is determined by action the discrete symmetry on $X_\mu$. Though we have shown a general coupling here, in what follows we assume vector-like couplings, or $C_A=0$.

\begin{figure}[t]
\begin{fmffile}{ABMix2}
\begin{fmfgraph*}(200,50)
\fmfpen{1.0}
\fmfstraight
\fmfleft{p1,i1,p2}\fmfright{p3,o1,p4}
\fmfv{l= $B_{ A}$}{i1}\fmfv{l=$B_B$}{o1}
\fmf{boson,tension=1.1}{i1,v1}
\fmf{zigzag, tension=0.9,label=$X$,label.side=left,label.dist=8}{v1,v2}
\fmf{boson, tension=1.1}{v2,o1}
\fmfv{decor.shape=circle,decor.filled=full,decor.size=1.5thick,l=$\varepsilon$,l.a=90,l.d=6}{v1}
\fmfv{decor.shape=circle,decor.filled=full,decor.size=1.5thick,l=$\varepsilon$,l.a=90,l.d=6}{v2}
\end{fmfgraph*}
\end{fmffile}
\caption{\label{f.ABMix2} Tree level mixing of SM and twin hypercharge bosons through their kinetic mixing with $X$.}
\end{figure}
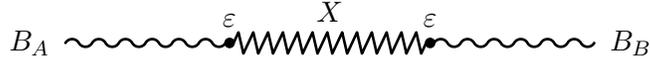

The bounds on the SM photon mixing directly by some parameter $\epsilon$ with a massless twin photon require $\epsilon\lesssim10^{-9}$~\cite{Davidson:2000hf,Vogel:2013raa}, for a twin electron with MeV scale mass. The most stringent result from bounds from supernova 1987A~\cite{Mohapatra:1990vq}, which applies to energy scales in the few to few hundred MeV range. 

With $X_\mu$ as intermediary, however, the two photons have tree level mixing at order $\varepsilon^2$, as shown in Fig.~\ref{f.ABMix2}. Thus, the supernova bound becomes $\varepsilon\lesssim10^{-4.5}$, which is too small a coupling to play a role in LHC searches. Even if the $X$ boson is massive, these bounds persist since the kinetic mixing operator can be generated by the running above $m_X$ if the sum of fermion charges is nonzero as discussed below.

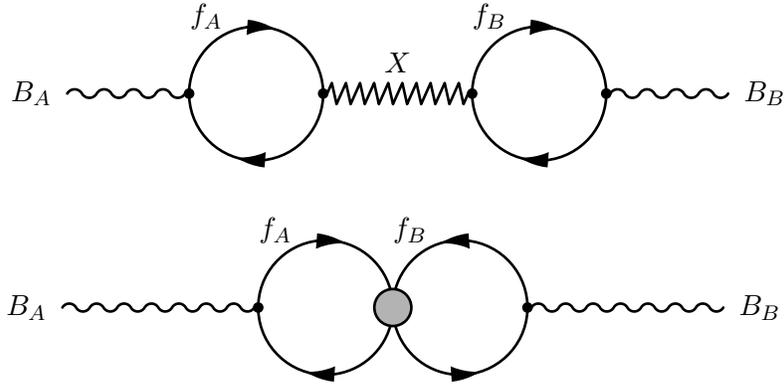
\begin{figure}[t]
\begin{fmffile}{ABMix}
\begin{fmfgraph*}(250,80)
\fmfpen{1.0}
\fmfstraight
\fmfleft{p1,i1,p2}\fmfright{p3,o1,p4}
\fmfv{l= $B_{ A}$}{i1}\fmfv{l=$B_B$}{o1}
\fmf{boson,tension=1.1}{i1,v1}
\fmf{fermion,tension=0.5,left=1}{v1,v2,v1}
\fmf{zigzag, tension=0.9,label=$X$,label.side=left,label.dist=8}{v2,v3}
\fmf{fermion,tension=0.5,left=1}{v3,v4,v3}
\fmf{boson, tension=1.1}{v4,o1}
\fmfv{decor.shape=circle,decor.filled=full,decor.size=1.5thick,l=$f_A$,l.a=75,l.d=25}{v1}
\fmfv{decor.shape=circle,decor.filled=full,decor.size=1.5thick}{v2}
\fmfv{decor.shape=circle,decor.filled=full,decor.size=1.5thick,l=$f_B$,l.a=75,l.d=25}{v3}
\fmfv{decor.shape=circle,decor.filled=full,decor.size=1.5thick}{v4}
\end{fmfgraph*}
\end{fmffile}\\
\begin{fmffile}{ABMixEft}
\begin{fmfgraph*}(250,80)
\fmfpen{1.0}
\fmfstraight
\fmfleft{p1,i1,p2}\fmfright{p3,o1,p4}
\fmfv{l= $B_{ A}$}{i1}\fmfv{l=$B_B$}{o1}
\fmf{boson,tension=1.1}{i1,v1}
\fmf{fermion,tension=0.8,left=1}{v1,v2,v1}
\fmf{fermion,tension=0.8,right=1}{v2,v4,v2}
\fmf{boson, tension=1.1}{v4,o1}
\fmfv{decor.shape=circle,decor.filled=full,decor.size=1.5thick,l=$f_A$,l.a=75,l.d=25}{v1}
\fmfv{decor.shape=circle,decor.filled=30,decor.size=7thick,l=$f_B$,l.a=75,l.d=25}{v2}
\fmfv{decor.shape=circle,decor.filled=full,decor.size=1.5thick}{v4}
\end{fmfgraph*}
\end{fmffile}
\caption{\label{f.ABMix} Loop level mixing of SM and twin hypercharge boson through $X$. Bottom graph with $X$ integrated out.}
\end{figure}

The kinetic mixing of $X$ with the other $U(1)$s from above the cutoff depends strongly on the field content, and so we take it as a free parameter $\varepsilon_\text{UV}$. This serves as a boundary condition for the running of the mixing operator
\begin{equation}
    B_{A\mu\nu}B^{\mu\nu}_B,
\end{equation}
which runs from the cutoff $\Lambda_\text{UV}$ down to $m_X$. The running can be calculated from the top diagram in Fig.~\ref{f.ABMix},
\begin{align}
    \varepsilon\sim& \frac{g_X^2g_Y^2}{576\pi^4} \left[\sum_{f}Y_{f}x_{f}\ln\frac{m_X^2}{\Lambda_\text{UV}^2}\right]^2.
\end{align}
Here the sum runs over all the fermions which couple to $X$ and $B$ and have masses below $m_X$. If some fermion has a mass above $m_X$, then it only contributes to the sum down to the mass of the fermion.

Taking the this sum to be order one, $g_X\sim g$, and $m_X$ a few hundred GeV we find $\varepsilon \sim 10^{-4.5}$. So, it seems we have already saturated the cosmological bound. But, it may be that
\begin{equation}
    \sum_{f}Y_{f}x_{f}=0.
\end{equation}
In which case there is no contribution from this running. A well known realization of this behavior is the gauged $L_\mu-L_\tau$ current. It is also anomaly free, and so does not require any new states at the cutoff, making a vanishing $\varepsilon_\text{UV}$ seem natural also. 

For vectorial $x_f$ couplings, the cancellation can also occur within each generation,
\begin{align}
    \sum_{f}Y_{f}x_{f}=&-2\frac12x_L+1\cdot x_\ell+ 3\cdot2\cdot\frac16x_Q-3\cdot\frac23x_u+3\cdot\frac13x_d\nonumber\\
    =&-2\frac12x_L-1\cdot x_L+ 3\cdot2\cdot\frac16x_Q +3\cdot\frac23x_Q -3\cdot\frac13x_Q\nonumber\\
    =&-2x_L+2x_Q=0.
\end{align}
But, this is only satisfied for $x_L=x_Q$, which makes $U(1)_X$ anomalous in the low energy theory. In particular, it violates the relation
\begin{equation}
    x_L+3x_Q=0,
\end{equation}
as required for the vanishing of the mixed $SU(2)_L^2U(1)_X$ anomaly. Resolving this anomaly requires new states charged under $SU(2)_L$ in the UV, which is difficult to reconcile with precision EW measurements.

We must now consider the running below $m_X$. In this case we integrate $X$ out of the spectrum and calculate by inserting the generated four-fermion operator in the lower diagram of Fig.~\ref{f.ABMix}. This yields
\begin{align}
    \sim&\frac{g_X^2g_Y^2}{576\pi^4}\frac{q^2}{m_X^2}\sum_{f_A}Y_{f_A}x_{f_A}\ln\frac{m_{f_A}^2}{m_X^2}\sum_{f_B}Y_{f_B}x_{f_B}\ln\frac{m_{f_B}^2}{m_X^2},
\end{align}
where again the sums over the $A$ and $B$ sector fermions include all the ``active" fermions over that mass range. The important thing to notice is the $q^2$ dependence. This diagram generates a higher dimensional operator, not the usual kinetic mixing.
    
This result can also be understood from the equations of motion. The fermion loops in Fig.~2 simply stand in for some kind of kinetic mixing $\hat{\varepsilon}$ between the $X$ and $B_{A,B}$, something of the form in Eq.~\eqref{e.kinMix}. The equations of motion for $X$ is then
\begin{equation}
    \left(\eta_{\mu\nu}\partial^2-\partial_\mu\partial_\nu+m_X^2\eta_{\mu\nu} \right)X^\nu=2\hat{\varepsilon}\partial^\nu\left(B_{A\nu\mu}\pm B_{B\nu\mu} \right)+\ldots\;,
\end{equation}
where we have not written the terms involving fermions. This can be inserted back into the Lagrangian to obtain terms like
\begin{equation}
    \frac{\hat{\varepsilon}^2}{m_X^2}B_{A,B\mu\nu}\partial^2B_{A,B}^{\mu\nu},
\end{equation}
which agrees with the explicit loop calculation.

Therefore, below $m_X$ the running of $\varepsilon$ can only occur at some higher loop order. If we assume vector-like couplings that have the same charge conjugation behavior as hypercharge, then only even numbers of $U(1)$ boson legs can connect to the fermsion loops. In this case the leading contribution to $\varepsilon$ would have to be at least four-loop, with three $X$ propagators integrated out. If such a diagram does generate kinetic mixing, it is likely well below the level of experimental constraint.

In short, it is consistent for $\varepsilon$ to be small enough to agree with cosmological constraints, but this greatly restricts the collider phenomenology. If, however, the twin hypercharge, $B_B$, also gets a mass $m_{B_B}\gtrsim1$ GeV, these bounds are much weaker~\cite{Essig:2013lka} and the kinetic mixing can be much larger. This gives both $X$ and both the massive neutral bosons in the twin sector $Z_B$ and $A_B$ large enough couplings to SM field to potentially produce them at colliders. The singleton $X$ can also keep the two sectors in thermal contact. 

In particular, it can put the decoupling temperature in between the QCD scales of the two sectors, reducing (though not eliminating) the tension with $\Delta N_\text{eff}$ bounds. The origin of $m_{B_B}$ is associated with a soft breaking of the $\mathbb{Z}_2$, the origin of which we take to lie above the low energy degrees of freedom. It could be arise from a St\"{u}ckelberg mechanism or from a scalar field getting a VEV, and spontaneously breaking the discrete symmetry and twin hypercharge. \footnote{The combination of lifting the twin neutrinos to a few GeV and lowering the decoupling temperature of the two sectors so that it is between the QCD phase transitions of either sector can reduce the MTH $\Delta N_\text{eff}$ to about 0.7, which may agrees with the BBN constraint which can be $\Delta N_\text{eff}\lesssim 1$~\cite{Cyburt:2004yc}. }

\subsubsection{Anomalous Singletons\label{ss.anom}}
The MTH construction only addresses the little hierarchy problem, its cutoff, $\Lambda_\text{UV}$, is a few TeV. Consequently, there is little obstruction to considering anomalous vector couplings. Such couplings merely require that new states appear below a scale~\cite{Preskill:1990fr}
\begin{equation}
    \Lambda_\text{Anomalous}\sim \frac{m_X}{g_X},
\end{equation}
to cancel the anomalies. As long as $\Lambda_\text{Anomalous}\geq\Lambda_\text{UV}$ there is no inconsistency. However, as shown below resolving the mixed anomalies in the UV requires particles charged under both $X$ and $B_{A,B}$. This generically generates kinetic mixing in the low energy EFT, so we find that anomalous currents lead to the kinetic mixing of Eq.~\eqref{e.kinMix}.

As a concrete example, suppose $X$ is the force carrier of gauged $L_\mu$, muon number. Then the $U(1)_X^3$ anomaly is proportional to
\begin{equation}
  \mathcal{A}_X\propto  C_\mu^3\pm C_\mu^3.\label{e.tripleAnom}
\end{equation}
When $X$ is odd under the discrete symmetry there is a relative sign between the terms and the anomaly vanishes. Both the $A$ and $B$ sector muons are crucial to the cancellation.

The mixed $U(1)_X$ anomalies are more subtle. For instance, the $U(1)_X\,U(1)_Y^2$ anomaly does not vanish. In the $A$ sector we have
\begin{equation}
\pbox[c]{\textwidth}{\includegraphics[scale=1]{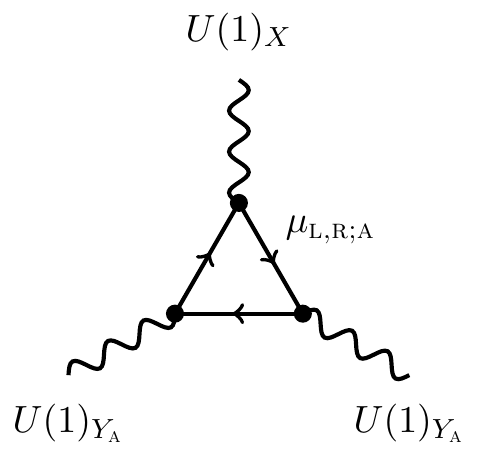}}\propto \left(Y_{\mu_L}^2-Y_{\mu_R}^2\right)x_{\mu_A} \neq 0.
\end{equation}
Which generates a term of the right hand side of the current conservation equation, 
\begin{equation}
  \left.  \partial_\mu J^\mu_X\right|_{Y_A}= \mathcal{A}_B\epsilon_{\mu\nu\alpha\beta}B_A^{\mu\nu}B_A^{\alpha\beta},
\end{equation}
where $\mathcal{A}_B$ is constant. By including the twin sector couplings we obtain the full equation
\begin{equation}
    \partial_\mu J^\mu_X= \mathcal{A}_Y\epsilon_{\mu\nu\alpha\beta}\left(B_A^{\mu\nu}B_A^{\alpha\beta}\pm B_B^{\mu\nu}B_B^{\alpha\beta} \right)+\mathcal{A}_X\epsilon_{\mu\nu\alpha\beta}X^{\mu\nu}X^{\alpha\beta}.
\end{equation}
While $\mathcal{A}_B$ is identical for the $A$ and $B$ sectors, the anomaly does not cancel, because each pertains to a distinct gauge field, either SM or twin hypercharge. In general, we cannot cancel $A$ sector anomalies by identical terms in the twin sector.

An anomalous $X_\mu$ has additional interactions among the gauge bosons~\cite{Antoniadis:2010zza} which can lead to stringent bounds on light vectors~\cite{Dror:2017ehi,Dror:2017nsg}. These follow from the Wess-Zumino terms in the low-energy EFT. 
For heavy vectors, these interactions are typically negligible. They introduce new decays like $X\to Z_{A,B}\gamma_{A,B}$, but the branching fractions are order $10^{-4}$~\cite{Ekstedt:2017tbo}. While these decays are very interesting and distinct, they are subdominant to other signatures, so we neglect them in what follows.

\section{Experimental Bounds on Vector Singletons \label{sec.model}}
In this section we determine the indirect and direct experimental bounds on vector singletons. We assume the usual MTH construction and add a new gauge group $U(1)_X$ with gauge boson $X_\mu$. These new gauge interactions may or may not be anomaly free. In particular, $(B-L)_\text{A}\pm(B-L)_\text{B}$ is used as an anomaly free case study, as well as the restriction to the third generation. We also consider the anomalous $L_{\mu_A}\pm L_{\mu_B}$ current. 

We take $X_\mu$ massive, through the Higgs or St\"{u}ckleberg mechanism. In general, we assume whatever dynamics gives mass to $X_\mu$ decouples from the low-energy dynamics and treat $m_X$ as a free parameter. The Lagrangian pertaining to $X_\mu$ is 
\begin{align}
\mathcal{L}_X=-\frac14 X_{\mu\nu}X^{\mu\nu}+\frac{m_X^2}{2}X_\mu X^\mu&+g_XX_\mu\left(J_A^\mu\pm J_B^\mu\right),\label{eq:lag-x}
\end{align}
where $J_i$ is the particular current of interest. Note that this Lagrangian is invariant under the $\mathbb{Z}_2$ symmetry if $X_\mu\to \pm X_\mu$. 

Appendix~\ref{a.kinmix} includes a perturbative form of the couplings and other formulae for these analyses. However, these results are only valid when $m_X$ is not too close to either $m_{Z_Z}$ or $m_{Z_B}$. For quantitative results we use a numerical diagonalization procedure, while the pertubative analysis provides qualitative understanding of those results. 

\subsection{Indirect Bounds}
\begin{figure}[t]
	\includegraphics[width=0.45\textwidth]{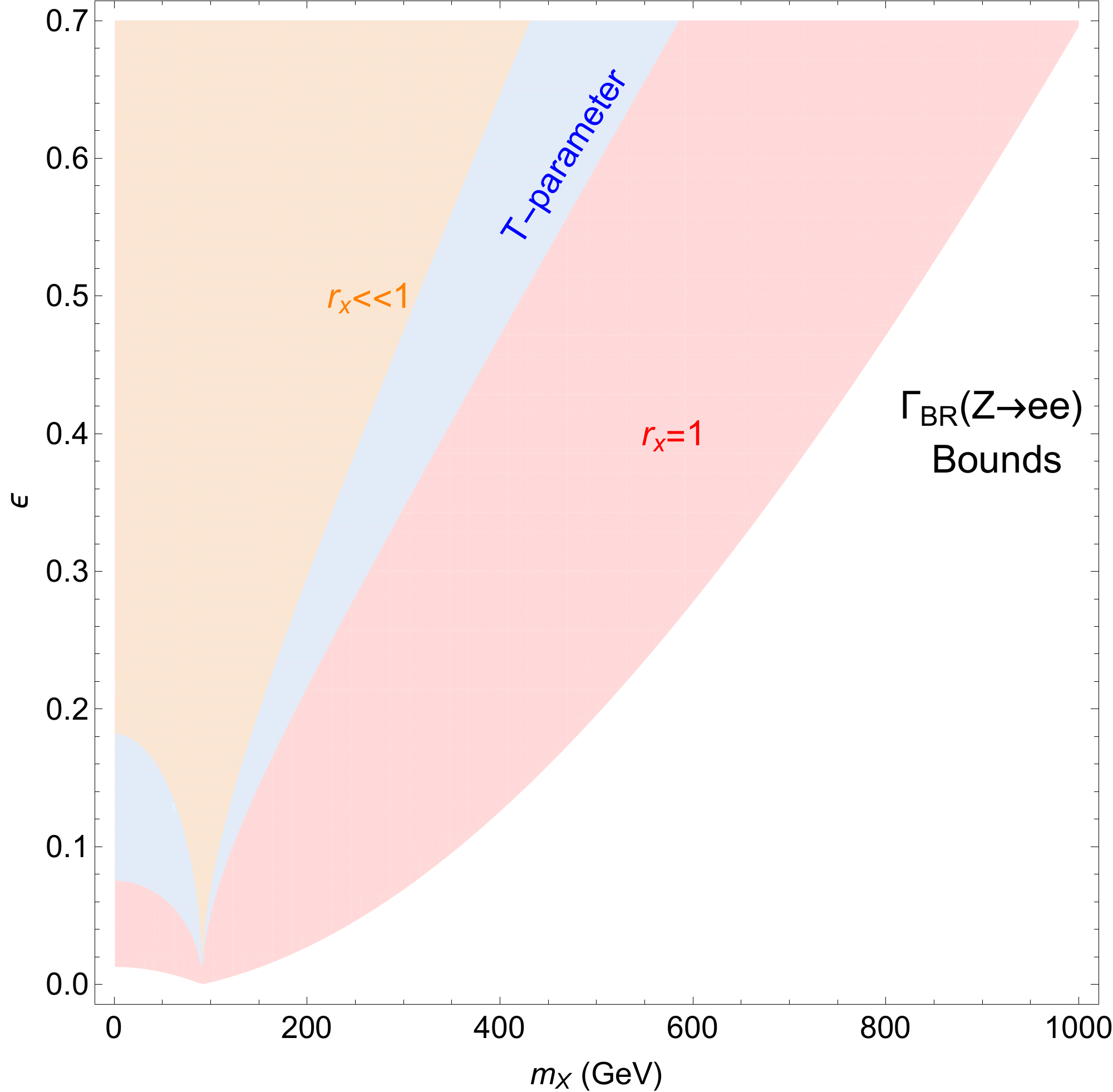}\hspace{4mm}
	\includegraphics[width=0.45\textwidth]{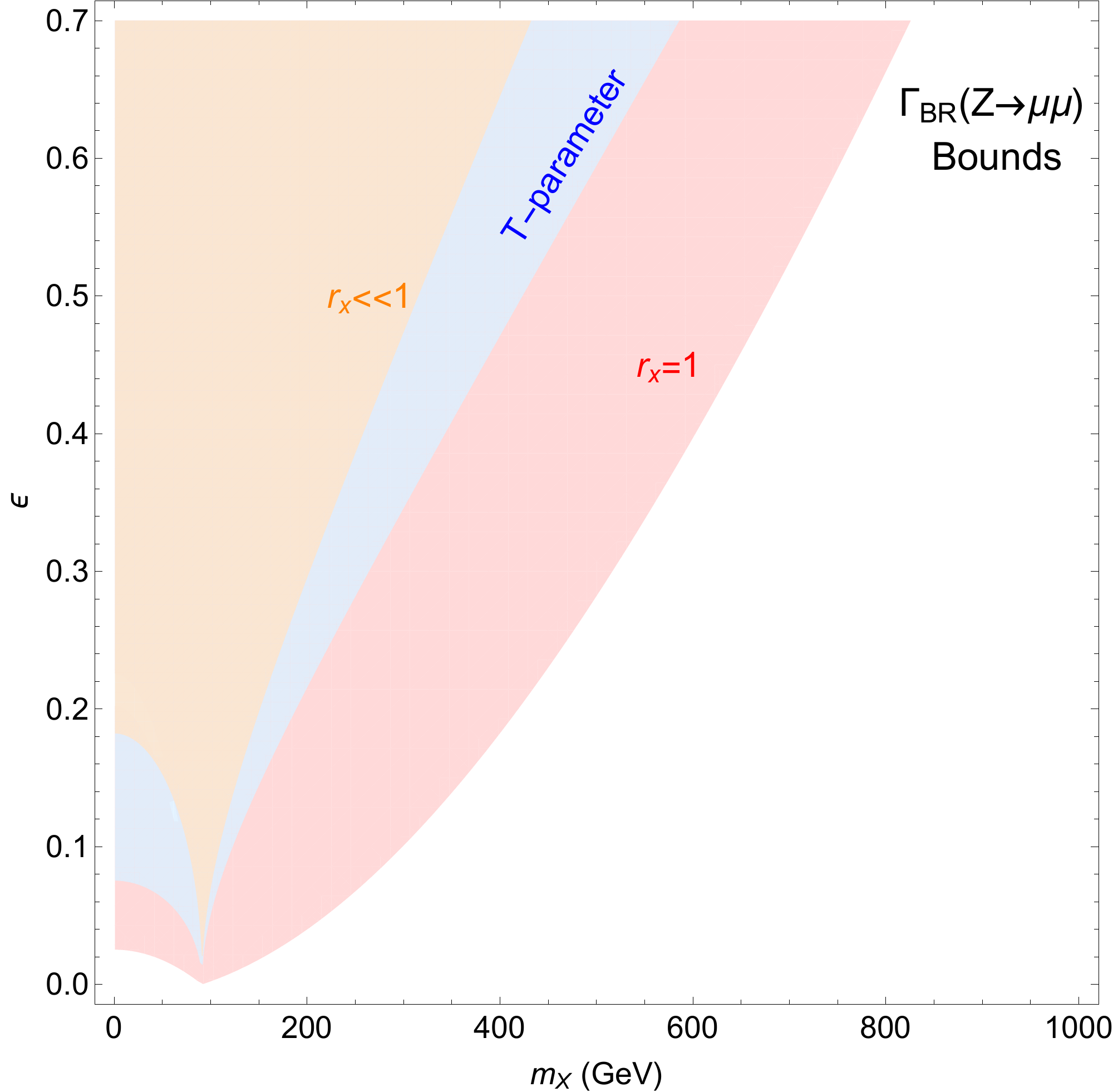}
	\caption{Bounds from precision $Z$ measurements. The orange (red) regions take $g_X\ll g$ ($g_X=g$), where $g$ is the SM weak gauge coupling. The red region assumes $X$ couples to $B-L$, giving a stronger bound. These two regions illustrate how strong or weak the bounds can be. The $T$-parameter bound also shown in blue.}
	\label{f.Zbounds}
\end{figure}

Indirect bounds on this scenario arise primarily from $Z$ physics at LEP. These include the $T$ parameter, the invisible $Z$ width, and the partial widths of the $Z$ into light leptons. As shown in Fig.~\ref{f.Zbounds}, these restrict the viable $(m_X,\varepsilon)$ space. 

From \cite{Appelquist:2002mw} the $T$ parameter satisfies
\begin{equation}
\frac{\alpha T}{1+\alpha T}=1-\frac{\widehat{m}_{Z_A}^2}{m_{Z_A}^2}\approx\varepsilon^2 s_W^2\frac{m_{Z_A}^2}{m_X^2}\frac{1}{1-m_{Z_A}^2/m_X^2},
\end{equation}
where $\alpha$ is the fine structure constant at the $Z$ mass, and in the last equality we have used the perturbative result in Eq.~\eqref{e.pertZmass}. From \cite{Baak:2014ora}  $T=0.09\pm0.13$ so, at 95\% confidence we need
\begin{equation}
-5.8\times 10^{-3}<\varepsilon^2 \frac{m_{Z_A}^2}{m_X^2}\frac{1}{1-m_{Z_A}^2/m_X^2}< 1.2\times10^{-2}.\label{e.Tparam}
\end{equation}
As the blue shaded region in Fig.~\ref{f.Zbounds} makes clear, this bound is most constraining for $m_X$ close to $m_{Z_A}$\footnotetext{Note that when the masses are degenerate a more careful treatment is required.}, and drops off steeply for larger masses. 

In this model, the $Z$ branching fractions into individual leptons are more constraining than its invisible width. We use the PDG~\cite{Tanabashi:2018oca} values $\mbox{BR}(Z\to ee)=(3.363\pm0.004)\%$ and $\mbox{BR}(Z\to \mu\mu)=(3.366\pm0.007)\%$ to determine 95\% confidence bounds. Unlike the $T$ parameter, these bounds are sensitive to the strength of the new gaugue coupling $g_X$. To indicate the range of the bounds, we use two benchmarks on $r_X=g_X/g$: $r_X\ll1$ and $r_X=1$. Here $g$ is the SM weak gauge coupling. The bounds are computed for the global $B-L$ case, as the larger number of direct couplings of $X_\mu$ to twin states changes the $Z_A$ width more, giving stronger bounds. The two regions, red and orange, in Fig.~\ref{f.Zbounds} correspond to these stronger and weaker bounds, respectively.

\subsection{LHC Bounds}

Both the singleton vector boson $X$ and the twin massive neutral gauge boson, $Z_B$, and in some cases $A_B$, are potentially accessible at the LHC. Since the $X$ can couple directly to $A$ and $B$ sector states, its couplings may not be suppressed by the kinetic mixing parameter $\varepsilon$. However, when it has no direct quark couplings, production arises through $\varepsilon$ suppressed mixing with the $Z_A$. In that case its production cross-section scales as $\varepsilon^2$.

On the other hand, $Z_B$ and $A_B$ only couples to SM quarks through kinetic mixing, and so their production at the LHC is always suppressed by at least $\varepsilon^2$. Their visible decays are similarly $\varepsilon^2$ suppressed. Therefore, the most promising discovery channel appears to be be jets + MET~\cite{Aaboud:2017buf}, thereby avoiding the extra suppression. However, even this least suppressed scenario gives sensitivity far below direct $X$ searches.

\subsubsection{Discovering the $X$}

The $X$ couplings depend on which combination of lepton and baryon numbers are gauged and, as discussed above, this significantly affects its signature at the the LHC. Here we explore the phenomenology of three gauged combinations of lepton and baryon numbers: $(B-L)_A-(B-L)_B$, $(B-L)_{3,A}-(B-L)_{3,B}$, and $L_{\mu_A}-L_{\mu_B}$. The first two cases can be made anomaly free if right-handed neutrinos are added to the SM field content while the last one requires additional fermions (see the discussion in Sec.~\ref{ss.anom}). The magnitude of the fermion $U(1)_X$ charges, $x_A$ and $x_B$, for these three cases are given in Tab.~\ref{tab:x-charges}.

\newcommand{\zero}{$\cdot$}
\begingroup
\setlength{\tabcolsep}{10pt}
\begin{table}[t]\centering
	\begin{tabular}{c c c c c c c c c c c}\toprule[1pt]
		Model & $x_A^q$ & $x_A^{t,b}$&$x_A^e$&$x_A^{\mu}$ &$x_A^\tau$ & $x_B^q$ & $x_B^{t,b}$&$x_B^e$&$x_B^{\mu}$  &$x_B^{\tau}$\\\midrule[0.5pt]
		$(B-L)_{A-B}$& ${}^{1}/{}_{3}$ & ${}^{1}/{}_{3}$ & 1 & 1 & 1 & ${}^{1}/{}_{3}$ & ${}^{1}/{}_{3}$ & 1 & 1 & 1\\
		$(B-L)_{3,A-3,B}$& \zero & ${}^{1}/{}_{3}$ & $\cdot$ & \zero & 1 & \zero & ${}^{1}/{}_{3}$ & \zero & \zero & 1\\
		$L_{\mu_A-\mu_B}$& \zero & \zero & \zero & 1 & \zero & \zero & \zero & \zero & 1 & \zero\\
	    \bottomrule[1pt]
	\end{tabular}  
	\caption{The magnitude of the $U(1)_X$ charges of $A$ and $B$ sector fermions for three cases $X$ currents. The superscripts specify the generation, with the first and second generation quarks denoted collectively by $q\in\{u,d,s,c\}$.}
	\label{tab:x-charges}
\end{table}
\endgroup

We first consider $X$ to be the gauge boson of generation-universal $(B-L)_A-(B-L)_B$. In this case $X$ couples directly to SM quarks and leptons, so the cross-section satisfies $\sigma\propto g_X^2\,x_A^2$. Here the charges, $x_A$, are given in Tab.~\ref{tab:x-charges} and the only free parameter (to leading order in $\varepsilon$) is $r_X=g_X/g$, i.e. the ratio of the $U(1)_X$ gauge coupling to the SM weak isospin gauge coupling. Note the sign of the $B$-sector charges is fixed in the Lagrangian Eq.~\eqref{eq:lag-x}.

\begin{figure}[t]
	\includegraphics[width=0.49\textwidth]{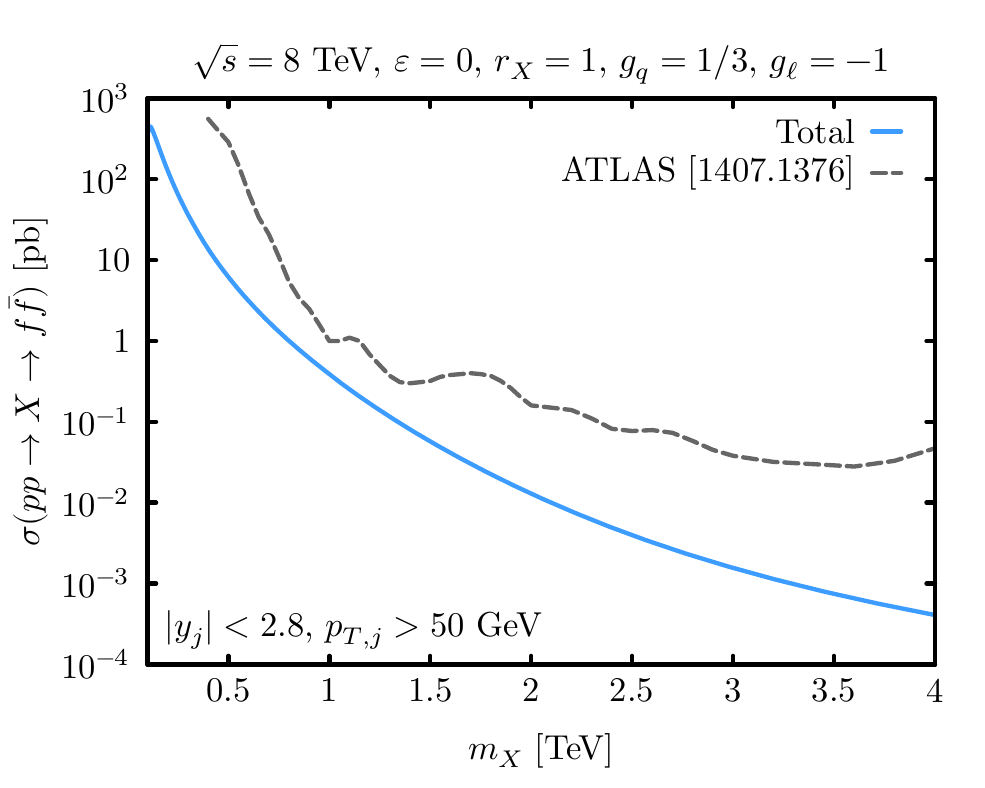}
	\includegraphics[width=0.49\textwidth]{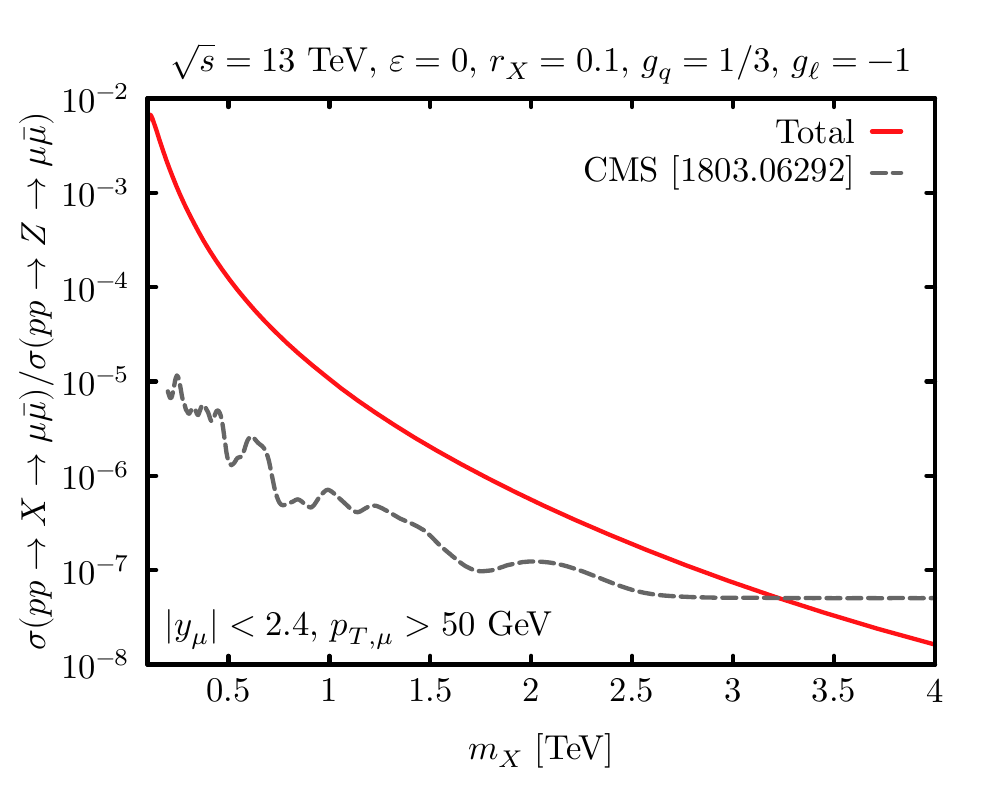}\\
	\includegraphics[width=0.49\textwidth]{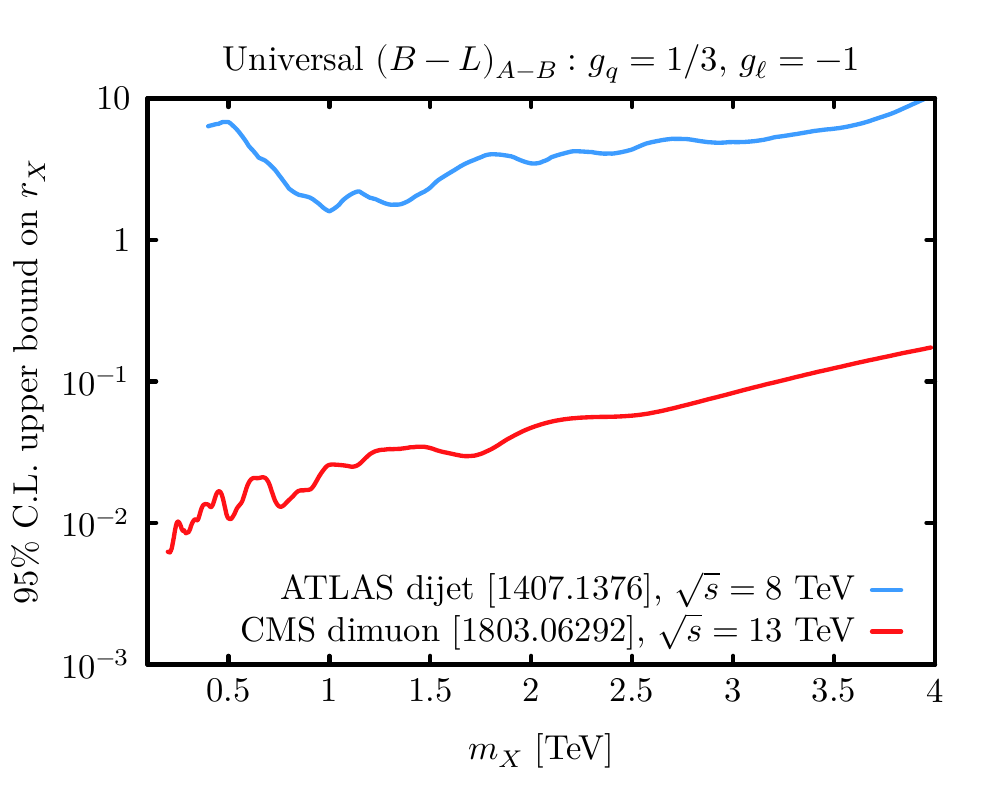}
	\caption{\textbf{Universal $\mathbf{(B-L)}$ model}. \emph{Top left panel:} the production cross-section at $\sqrt{s}=8$ TeV and the 95\% C.L. limit from the ATLAS dijet resonance search~\cite{Aad:2014aqa}. \emph{Top right panel:} The ratio of the production cross-section to the SM $Z$ cross-section at $\sqrt{s}=13$ TeV and the 95\% C.L. limit from the CMS dilepton search in the dimuon channel~\cite{Sirunyan:2018exx}. \emph{Bottom panel:} the recast ATLAS and CMS limits on $r_X$. 
	}
	\label{fig:dijet-and-dilep-bml}
\end{figure}

The size of $r_X$ is constrained by dijet and dilepton resonances searches at the LHC. For example, Fig.~\ref{fig:dijet-and-dilep-bml} shows the bound from an ATLAS dijet~\cite{Aad:2014aqa} (top left panel) and CMS dilepton~\cite{Sirunyan:2018exx} (top right panel) searches along with the predicted fiducial cross-section for $r_X=1$. 
The bound these place on $r_X$ is shown in the bottom panel of Fig.~\ref{fig:dijet-and-dilep-bml}. In particular, the di-muon channel of the CMS dilepton search yields a bound of $r_X\sim 10^{-2}-10^{-1}$ up to multi-TeV masses.

\begin{figure}[t]
	\includegraphics[width=0.49\textwidth]{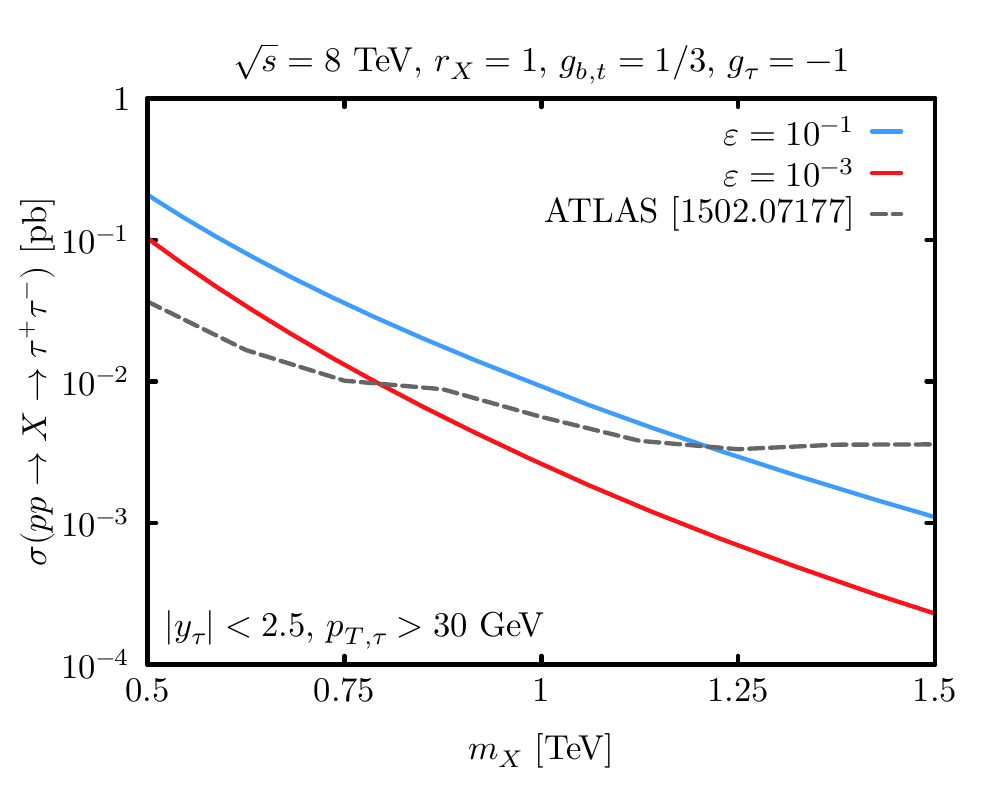}
	\includegraphics[width=0.49\textwidth]{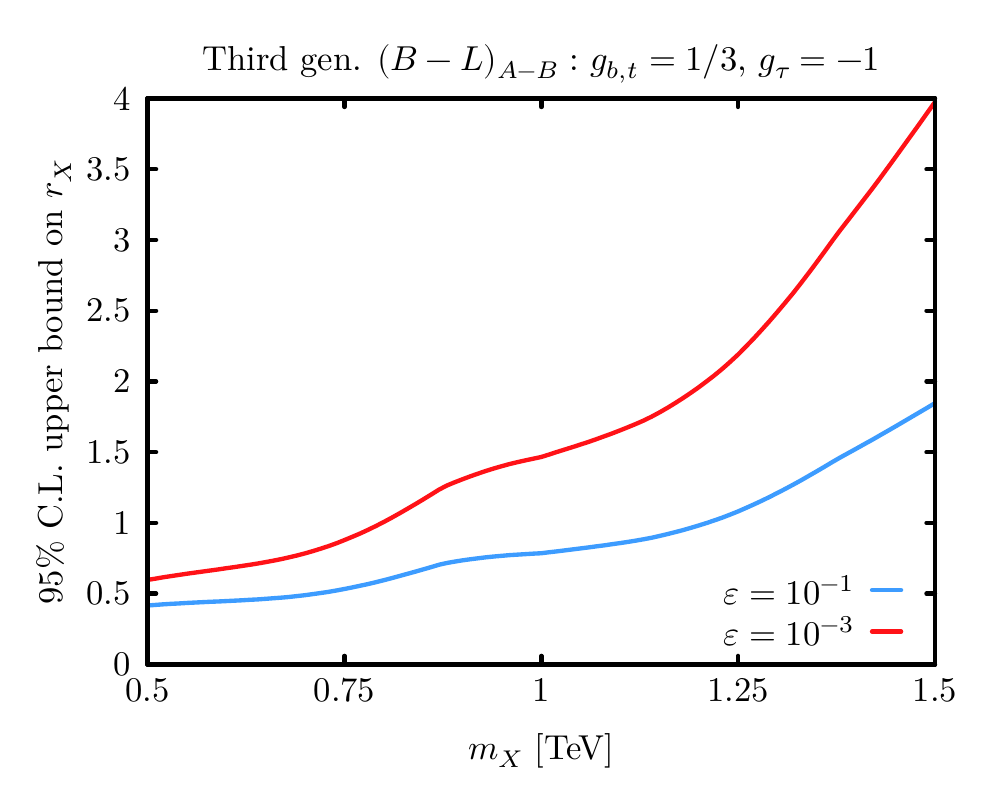}
	\caption{\textbf{3\textsuperscript{rd} gen. $\mathbf{(B-L)}$ model}. \emph{Left panel:} the production cross-section at $\sqrt{s}=8$ TeV and the 95\% C.L. limit from the ATLAS $\tau^+\tau^-$ resonance search~\cite{Aaboud:2017buf}. \emph{Right panel:} the recast ATLAS limit for $\varepsilon=10^{-3}$ and $10^{-1}$. For large values of $\varepsilon$, the production cross-section is enhanced due to light-quark contributions.
	}
	\label{fig:ditau-b3ml3}
\end{figure}

In the second case of interest we gauge the difference of $A$ and $B$ sector $B-L$ currents of the third generation only. Consequently, production at the LHC proceeds either through the $b\bar{b}$ initial state, with the corresponding PDF suppression, or through light quarks with $\varepsilon^2$ suppression. The latter is suppressed because the couplings to light quarks are only generated by mixing with $Z_A$. The dominant final states in this case are $b\bar{b}$, $t\bar{t}$, and $\tau^+\tau^-$ all of which are covered by current searches, see for example~\cite{Aaboud:2018tqo,Aaboud:2018mjh,Khachatryan:2016qkc}. Currently, the strongest bound is from $\tau^+\tau^-$ searches and is shown in Fig.~\ref{fig:ditau-b3ml3}. The plot on the right shows the bounds on $r_X$ as a function of $m_X$ for two $\varepsilon$ benchmark values. We see that for $m_X\sim$500 GeV the new gauge coupling cannot be more than half the SM weak-isospin gauge coupling, but this bound relaxes as the $X$ mass increases.

\begin{figure}[t]
	\includegraphics[width=0.49\textwidth]{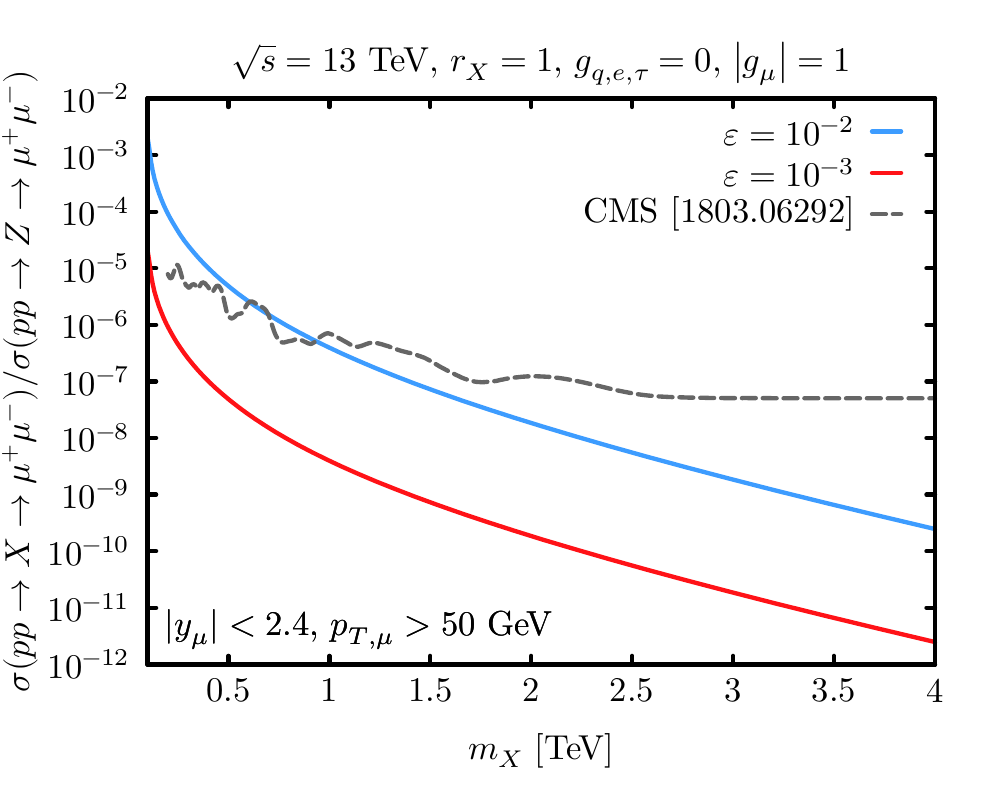}
	\includegraphics[width=0.49\textwidth]{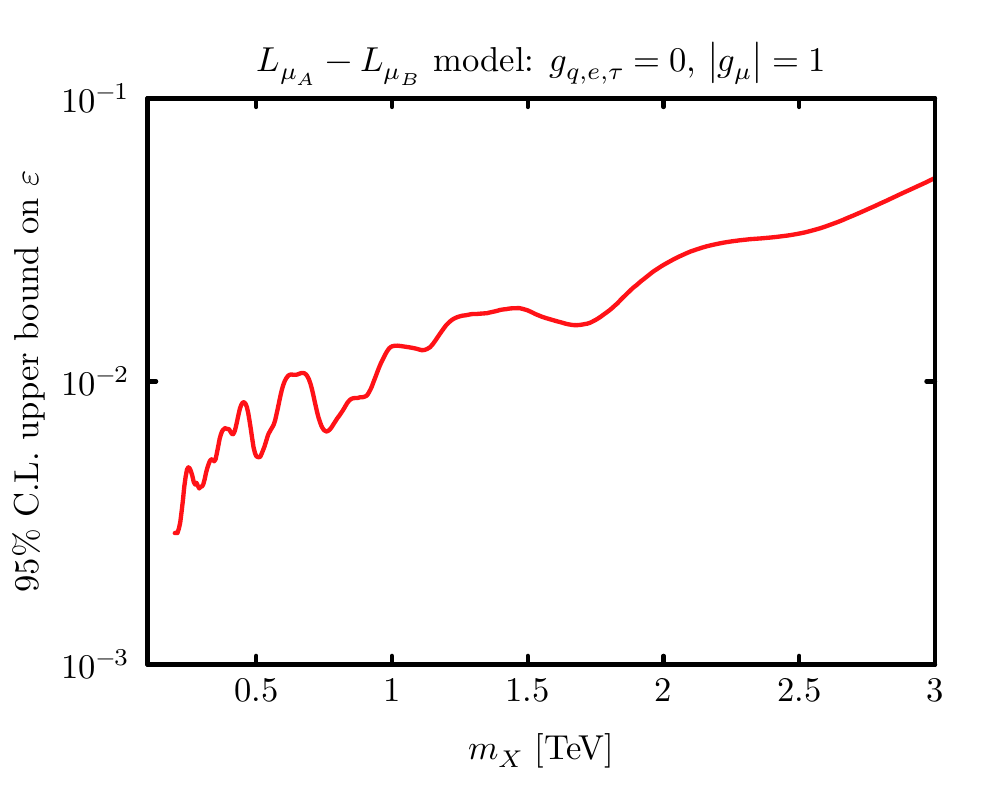}
	\caption{\textbf{$\mathbf{L_{\mu_A}-L_{\mu_B}}$ model.} \emph{Left panel}: the ratio of the production cross-section to the SM $Z$ cross-section as a function of $m_X$ and the 95\% C.L. limit from the CMS dilepton resonance search at 13 TeV~\cite{Sirunyan:2018exx} for the dimuon final state. \emph{Right panel:} Recast of the CMS limit on the kinetic mixing parameter $\varepsilon$.}
	\label{fig:dimuon-mua-mub}
\end{figure}

Finally, we gauge the difference of $A$ and $B$ sector muon numbers. Now production must proceed via order $\varepsilon$ couplings to light quarks, which are generated through mixing with $Z_A$. As a result, the cross-section is suppressed by $\varepsilon^2$. Further, because only the dimuon final state carries no additional powers $\varepsilon$, the strongest bound comes from dimuon searches. Fixing $r_X=1$, the bound on $\varepsilon$ is shown in Fig.~\ref{fig:dimuon-mua-mub} and is $\mathcal{O}(10^{-2})$.

\setcounter{topnumber}{1}

\subsubsection{Discovering the twin Z and A}
The kinetic mixing mediated by $X$ leads SM quarks and leptons to obtain small couplings to the neutral bosons in the twin sector. The mass of $Z_B$ is controlled by ratio of Higgs VEVs in the two sectors, $v_B/v_A$. This ratio is also tied to limits on deviations in Higgs couplings~\cite{Burdman:2014zta} and imply $m_{Z_B}\gtrsim 3 m_{Z_A}$, while requiring the model be no more than 10\% tuned ensures $m_{Z_B}\lesssim 6 m_{Z_A}$. The $A_B$ mass is much less constrained, as long as it is at least a few GeV to agree with cosmological bounds. This means that both neutral twin gauge bosons can be considerably lighter than $X$.

The same search strategies used to detect $X$ apply to $Z_{B}$ and $A_B$, with a few important differences. Because $Z_B$ and $A_B$ couple to $A$ sector fields through mixing with $X$, their couplings are $\varepsilon$ suppressed relative to the $X$ coupling. From Eqs.~\eqref{e.Aagff}\textendash\eqref{e.Xgff} in App.~\ref{a.kinmix}, which apply in the massless twin photon limit,
\begin{equation}
    g_{Z_Bf_Af_A}\sim\varepsilon g_{X f_A f_A}.
\end{equation}
In addition, when $X$ is heavier than $Z_B$ this mixing further reduced, 
\begin{equation}
    g_{Z_Bf_Af_A}\sim\varepsilon \frac{m_{Z_B}^2}{m_X^2}g_{X f_A f_A}. \ \ \ \ m_X>m_{Z_B}.
\end{equation}
This same behavior is inherited by the massive twin photon, since its couplings come from an additional mass mixing with the twin $Z$.

\subsubsection{Testing the Twin Higgs Structure}
Discovering the $X$ would give us access to a new sector but it may be difficult to distinguish a twin sector from something generic. A high energy lepton collider might measure the $X$ width, and show that full width is twice the visible width. This would at least hint at the twin structure. Otherwise, the $X$ introduces many new parameters, and so its discovery does not probe the constrained nature of the twin Higgs framework.

Discovering both $X$ and a twin boson, say $Z_B$ does not significantly improve the prospects of testing the model. This is because the $X$ couplings depend on $g_X$ and the leading couplings to $Z_B$ depend on $g_X \epsilon$. Consequently the production rates of the two states are related by a free parameter. 

However, discovering both $Z_B$ and $A_B$ is different. Finding both states gives access to four experimental results, the two masses $m_{Z_B}$ and $m_{A_B}$ and the rates into the final states used to discover them. Both of these rates depend on the combination $g_X\varepsilon$ in the same way. In this case much of the dependence on $X$ factors out.

The set of unknown parameters of the model thus are: $f$, $g_X$, $m_X$, $m_{A_B}$, $\varepsilon$, and the set of $x_f$. If the discoveries occur in the same channel, di-leptons say, the combination $g_X\varepsilon x_f$ will be common to the two rates. Furthermore, the dependence on each vector's total width on $m_X$ is higher order in $\varepsilon$ -- see Eqs.~\eqref{e.Abgff} and \eqref{e.Zbgff} -- and can thus be neglected to leading order. The only remaining uncertainty is that $X$ might couple directly to other states, for instance if the discovery were made in di-leptons the $X$ might also couple to quarks directly.

This uncertainly could likely be reduced by checking other search channels directly at the $m_{Z_B}$ and $m_{A_B}$ masses. These additional excesses would fill out the number of $X$ couplings. If no additional excess is seen in those channels, the corresponding $x_f$ charges can be bounded. Then the twin Higgs structure would make a precise prediction of the widths of each vector. The measured rate of one vector could then be compared with the other, testing the twin Higgs structure.

\section{Conclusion\label{sec:conclusion}}
The mirror twin Higgs has become one of the most natural ways to resolve the hierarchy problem in the wake of LHC searches for colored symmetry partners. However, the minimal form of the twin Higgs along with the discrete symmetry seem to make connecting the two sectors, and hence probing the new states, difficult. In this work we have introduced a new type of portal between the SM and twin sectors, which has no twin under the discrete symmetry, transforming by at most a phase. These singleton portals provide new opportunities to probe the twin sector and may provide the means to produce novel phenomena within the twin Higgs framework.

At the renormalizable level, scalar singletons primarily affect the Higgs potential. Though they can easily explain the origin of soft $\mathbb{Z}_2$ breaking, their collider signals may be beyond current experimental capabilities. However, their mixing with the radial mode of the symmetry breaking, the heavy twin Higgs, could well be stronger and could lead to measurable effects.

Right-handed neutrinos are interesting fermion singletons. However, the variety of constraints on neutrino physics limits how they connect the sectors. Specific choices of couplings in flavor space or inverse seesaw like-constructions are consistent with existing constraints, and may be probed by upcoming examination of the $\beta$-decay spectrum. Further investigation of these twin neutrino interactions is certainly warranted.

Vector singletons provide a varied class of portals. They can have appreciable couplings to both sectors and still remain consistent with present experimental results. Such vectors may play a role in explaining cosmological data or the persistent anomalies in the decays of heavy flavor mesons. Further study of their explanatory power and phenomenology is both motivated, and ongoing.

These vectors are discoverable at the LHC and future colliders. These next generation machines may also be able to use the singleton set-up to probe the twin sector. They may even be able to distinguish the twin Higgs framework from more generic new physics.

\acknowledgments
We are grateful to Zackaria Chacko for many valuable comments and advice regarding the manuscript, as well as Adam Falkowski, Markus Luty, and John March-Russell for helpful discussions. We also thank the organizers and participants of the Flavour and Dark Matter 2017 workshop where this work began. C.B.V. is supported
by Department of Energy DE-SC-000999 and also thanks the Rudolf Peierls Centre for Theoretical Physics for hospitality during the completion of this work. This work was performed in part at Aspen Center for Physics, which is supported
by National Science Foundation grant PHY-1607611. This work was partially supported by a grant from the Simons Foundation. 

\appendix

\section{Kinetic mixing and Couplings}
\label{a.kinmix}

In this Appendix we calculate the masses and couplings of the massive vectors to leading order in the kinetic mixing parameter $\varepsilon$. We begin by writing the total kinetic mixing in the Lagrangian as 
\begin{equation}
-\frac14\left(\begin{array}{ccc}
B_{A\mu\nu} & B_{B\mu\nu} & X_{\mu\nu}
\end{array} \right)\left( \begin{array}{ccc}
1 & 0 & -\varepsilon\\
0 & 1 &\mp \varepsilon\\
-\varepsilon & \mp\varepsilon & 1
\end{array}\right)\left(\begin{array}{c}
B_{A\mu\nu} \\
B_{B\mu\nu} \\
X_{\mu\nu}
\end{array} \right).
\end{equation}
Then, by making the field redefinitions
\begin{equation}
\left(\begin{array}{c}
B_{A\mu} \\
B_{B\mu} \\
X_{\mu}
\end{array} \right)=\left( \begin{array}{ccc}
1+f(\varepsilon) & \pm f(\varepsilon) & 0\\
\pm f(\varepsilon) & 1+f(\varepsilon) & 0\\
\varepsilon(1+2f(\varepsilon)) & \pm\varepsilon(1+2f(\varepsilon)) & 1
\end{array}\right)\left(\begin{array}{c}
\overline{B}_{A\mu} \\
\overline{B}_{B\mu} \\
\overline{X}_{\mu}
\end{array} \right),
\end{equation}
with 
\begin{equation}
f(\varepsilon)=-\frac12 +\frac{1}{2\sqrt{1-2\varepsilon^2}}.
\end{equation}
the kinetic terms are diagonalized. Note, however, that this parameterization is only valid for $\varepsilon\leq 1\sqrt{2}$.\footnote{This bound on $\varepsilon$ is simply the requirement that there be not ghosts, as can arise if the mixing is too large. It is straightforward to prove that if $X_\mu$ had mixed with $N$ separate gauge bosons then only $\varepsilon\leq 1/\sqrt{N}$ keeps keeps the determinant of the mixing matrix positive.}

Now, the photon $A_\mu$ and neutral weak boson $Z_\mu$ are defined by
\begin{equation}
A_{A,B\mu}=c_WB_{A,B\mu}+s_WW^3_{A,B\mu}, \ \ \ \ Z_{A,B\mu}=-s_WB_{A,B\mu}+c_WW^3_{A,B\mu},
\end{equation}
where $c_W\equiv\cos\theta_W$ and similar for $s_W$. We then make the definitions
\begin{equation}
\overline{A}_{A,B\mu}\equiv c_W\overline{B}_{A,B\mu}+s_WW^3_{A,B\mu}, \ \ \ \ \overline{Z}_{A,B\mu}\equiv-s_W\overline{B}_{A,B\mu}+c_WW^3_{A,B\mu},
\end{equation}
which lead to mass terms 
\begin{align}
&\frac{m^2_{Z_A}}{2}\left\{ \overline{Z}_{A}^\mu\left(1+s_W^2f\right)
-s_Wf\left[c_W\left(\overline{A}_A^\mu\pm\overline{A}_B^\mu \right)\mp s_W\overline{Z}_B^\mu \right]\right\}^2\nonumber\\
&+\frac{m^2_{Z_B}}{2}\left\{ \overline{Z}_{B}^\mu\left(1+s_W^2f \right)
\mp s_Wf\left[c_W\left(\overline{A}_A^\mu\pm\overline{A}_B^\mu \right)- s_W\overline{Z}_A^\mu \right]\right\}^2\nonumber\\
&+\frac{m_X^2}{2}\left\{\overline{X}^\mu+\varepsilon(1+2 f) \left[c_W\left(\overline{A}_A^\mu\pm\overline{A}_B^\mu \right)-s_W\left(\overline{Z}_A^\mu\pm\overline{Z}_B^\mu \right) \right] \right\}^2\nonumber\\
&+\frac{m_{B_B}^2}{2}\left\{c_W\overline{{A}}_B^\mu-s_W\overline{Z}_B^\mu+f\left[c_W\left(\overline{A}_B^\mu\pm\overline{A}_A^\mu \right)-s_W\left(\overline{Z}_B^\mu\pm\overline{Z}_A^\mu \right) \right] \right\}^2.
\end{align}

To give the twin photon a mass, the $2\times 2$ mass matrix of $(Z_B,A_B)$ must be rank 2. To do this, we add another mass term for the hypercharge gauge boson $B_B$ of the form
\begin{equation}
\Delta m_{B_B}^2=\frac{1}{2}\frac{v_B^2}{4}\,g_2^2\,\alpha^2\,.
\end{equation}

Finally, we diagonalize the mass matrix to determine the physical eigenstates. For simplicity we give the results with $\Delta m^2_{B_B}=0$ as they illustrate the qualities pointed to the in the main text. The eigen vales of this matrix can be obtained in powers of $\varepsilon$:
\begin{align}
\widehat{m}_X^2=&m_X^2+\varepsilon^2m_X^2\left[2c_W^2 +s_W^2\left(M_{XZ_A}+M_{XZ_B}\right) \right]+\mathcal{O}(\varepsilon^4),\\
\widehat{m}_{Z_A}^2=&m_{Z_A}^2+\varepsilon^2s_W^2m_{Z_A}^2M_{Z_AX}+\mathcal{O}(\varepsilon^4),\label{e.pertZmass}\\
\widehat{m}_{Z_B}^2=&m_{Z_B}^2+\varepsilon^2s_W^2m_{Z_B}^2M_{Z_BX}+\mathcal{O}(\varepsilon^4),
\end{align}
while the $\widehat{A}_{A,B\mu}$ fields are massless. Here we have used the notation
\begin{equation}
    M_{AB}\equiv\frac{m_A^2}{m_A^2-m_B^2}.
\end{equation}
Note that when the masses are close to degenerate the perturbative expansion breaks down. 

We find the mass eigenstates are related to the interaction fields by

\begin{align}
A_{A\mu}=&\widehat{A}_{A\mu}-\varepsilon^2c_Ws_WM_{Z_AX}\widehat{Z}_{A\mu} \mp\varepsilon^2\widehat{A}_{B\mu} \mp\varepsilon^2c_Ws_WM_{Z_BX}\widehat{Z}_{B\mu}+\varepsilon c_W\widehat{X}_\mu,\label{e.Aadef}\\
Z_{A\mu}=&\left[ 1+\frac{\varepsilon^2}{2}s_W^2M_{Z_AX}\left(M_{Z_AX}+2M_{XZ_A} \right)\right]\widehat{Z}_{A\mu} \pm\varepsilon^2s_W^2M_{Z_BX}M_{Z_BZ_A}\widehat{Z}_{B\mu}-\varepsilon s_WM_{XZ_A}\widehat{X}_\mu,\label{e.Zadef}\\
A_{B\mu}=&\widehat{A}_{B\mu}-\varepsilon^2c_Ws_WM_{Z_BX}\widehat{Z}_{B\mu}\pm\varepsilon^2\widehat{A}_{A\mu}\mp\varepsilon^2c_Ws_WM_{Z_AX}\widehat{Z}_{A\mu}\pm\varepsilon c_W\widehat{X}_\mu,\label{e.Abdef}\\
Z_{B\mu}=&\left[ 1+\frac{\varepsilon^2}{2}s_W^2M_{Z_BX}\left(M_{Z_BX}+2M_{XZ_B}\right)\right]\widehat{Z}_{B\mu}\pm\varepsilon^2s_W^2M_{Z_AX}M_{Z_AZ_B}\widehat{Z}_{A\mu}\mp\varepsilon s_WM_{XZ_B}\widehat{X}_\mu,\label{e.Zbdef}\\
X_{\mu}=&\left\{1+\left[1-\frac{s_W^2}{2}\left(M_{Z_AX}^2+M_{Z_BX}^2\right) \right]\varepsilon^2 \right\}\widehat{X}_\mu-\varepsilon s_WM_{Z_AX}\widehat{Z}_{A\mu}\mp\varepsilon s_WM_{Z_BX}\widehat{Z}_{B\mu}\label{e.Xdef}
\end{align}

To determine couplings we consider the covariant derivative, neglecting gluon and $W^\pm$ boson fields. In doing so we define the electric charge of a fermion as $Q_A$ and its twin electric charge as $Q_B$. So, the electron has $Q_A=-1$ and $Q_B=0$. The $SU(2)$ charges $T_{3A}$ and $T_{3B}$ are similarly defined with, for example, the twin left-handed top quark having $T_{3A}=0$ and $T_{3B}=\tfrac12$. Finally, a fermion's $X_\mu$ charge is $x_{A,B}$.
\begin{align}
D_\mu=&\partial_\mu+ieQ_AA_{A\mu}+i\frac{g}{c_W}(T_{3A}-s_W^2Q_A)Z_{A\mu}\nonumber\\
&+ieQ_BA_{B\mu}+i\frac{g}{c_W}(T_{3B}-s_W^2Q_B)Z_{B\mu}+ig_X(x_A+x_B)X_\mu,
\label{eq:covder-gauge-basis}
\end{align}

Then from the definitions of the physical eigenstates in Eqs.~\eqref{e.Aadef}\textendash\eqref{e.Xdef} we find
\begin{align}
g_{\widehat{A}_Aff}=&e\left(Q_{A}\pm\varepsilon^2Q_{B}\right),\label{e.Aagff}\\
g_{\widehat{Z}_Aff}=&\frac{g}{c_W}\left\{\left(T_{3A}-s_W^2Q_{A} \right)\left[1+\frac{\varepsilon^2}{2}s_W^2M_{Z_AX}\left(M_{Z_AX}+2M_{XZ_A} \right) \right]-\varepsilon^2\left(Q_A\pm Q_B\right)s_W^2c_W^2M_{Z_AX}\right.\nonumber\\
&\left.\phantom{\frac{g}{c_W}}-\varepsilon r_X(x_A+x_B)s_Wc_WM_{Z_AX}
\pm\varepsilon^2\left(T_{3B}-s_W^2Q_B \right)s_W^2M_{Z_AX}M_{Z_AZ_B}\right\},\label{e.Zagff}\\
g_{\widehat{A}_Bff}=&e\left(Q_{B}\mp\varepsilon^2Q_{A}\right),\label{e.Abgff}\\
g_{\widehat{Z}_Bff}=&\frac{g}{c_W}\left\{\left(T_{3B}-s_W^2Q_{B} \right)\left[1+\frac{\varepsilon^2}{2}s_W^2M_{Z_BX}\left(M_{Z_BX}+2M_{XZ_B} \right) \right]-\varepsilon^2\left(Q_B\pm Q_A\right) s_W^2c_W^2M_{Z_BX}\right.\nonumber\\
&\left.\phantom{\frac{g}{c_W}}\mp\varepsilon r_X(x_A+x_B)s_Wc_WM_{Z_BX}
\pm\varepsilon^2s_W^2\left(T_{3A}-s_W^2Q_A \right)M_{Z_BX}M_{Z_BZ_A} \right\},\label{e.Zbgff}\\
g_{\widehat{X}ff}=&\frac{g}{c_W}\left\{c_Wr_X (x_A+x_B)\left(1+\varepsilon^2\left[1-\frac{s_W^2}{2}\left(M_{Z_AX}^2+M_{Z_BX}^2\right) \right] \right)+\varepsilon \left(Q_A\pm Q_B\right)s_Wc_W^2\right.\nonumber\\
&\left.\phantom{\frac{g}{c_W}}-\varepsilon s_W\left[\left(T_{3A}-s_W^2Q_A \right)M_{XZ_A}
\pm\left(T_{3B}-s_W^2Q_B \right)M_{XZ_B}\right]\right\},\label{e.Xgff}
\end{align}
where $r_X\equiv g_X/g$. Note that the electric charges are unchanged, as dictated by gauge invariance. However, the SM particles do obtain charges under the twin EW gauge fields, and the coupling of the SM fermions to the $Z$ is also changed.  

The coupling of the vectors to SM and twin $W$s are
\begin{align}
g_{\widehat{A}_AW_AW_A}=& e, &g_{\widehat{A}_AW_BW_B}=&\pm e\varepsilon^2,\\
g_{\widehat{A}_BW_AW_A}=&\pm e\varepsilon^2 ,&g_{\widehat{A}_BW_BW_B}=&e,\\
g_{\widehat{Z}_AW_AW_A}=& gc_W\left[1-\frac{\varepsilon^2}{2}s_W^2M_{Z_AX}^2 \right], &g_{\widehat{Z}_AW_BW_B}=&\mp gc_W\varepsilon^2s_W^2M_{Z_AX}M_{Z_BX},\\
g_{\widehat{Z}_BW_AW_A}=&\pm gc_W\varepsilon^2s_W^2M_{Z_AX}M_{Z_BX} ,&g_{\widehat{Z}_BW_BW_B}=&gc_W\left[1-\frac{\varepsilon^2}{2}s_W^2M_{Z_BX}^2 \right],\\
g_{\widehat{X}W_AW_A}=&\varepsilon gc_Ws_WM_{Z_AX}, & g_{\widehat{X}W_BW_B}=&\pm\varepsilon gc_Ws_WM_{Z_BX}.
\end{align}

The couplings to the SM Higgs $h$ and a second vector spring from the interactions
\begin{equation}
\frac{m_{Z_A}^2}{v_\text{EW}}\cos\vartheta hZ_{A\mu}Z_A^\mu, \ \ \ \ \frac{m_{Z_B}^2}{v_\text{EW}}\tan\vartheta\sin\vartheta hZ_{B\mu}Z_B^\mu,
\end{equation}
where $\vartheta=v/(f\sqrt{2})$ where $f$ is the VEV the breaks the global $SU(4)$ symmetry in the Higgs sector and $v$ is the amount of the VEV that is in the SM sector. In this parameterization $v_\text{EW}=f\sqrt{2}\sin\vartheta$. It is also worth noting that $m_{Z_B}=m_{Z_A}\cot\vartheta$, so that the couplings of the Higgs to $Z$ bosons in each sector are equal, as expected from the gauge structure. The couplings are
\begin{align}
g_{h\widehat{Z}_A\widehat{Z}_A}=&\frac{m_{Z_A}^2}{v_\text{EW}}\cos\vartheta\left(1+\varepsilon^2s_W^2M_{Z_AX}^2 \right), \\
g_{h\widehat{Z}_A\widehat{X}}=&-2\varepsilon\frac{m_{Z_A}^2}{v_\text{EW}}\cos\vartheta s_WM_{XZ_A},\\
g_{h\widehat{Z}_B\widehat{Z}_B}=&=\frac{m_{Z_A}^2}{v_\text{EW}}\cos\vartheta\left[1+\varepsilon^2s_W^2M_{Z_BX}^2 \right], \\
g_{h\widehat{Z}_B\widehat{X}}=&\mp 2\varepsilon\frac{m_{Z_A}^2}{v_\text{EW}}\cos\vartheta s_WM_{XZ_B},\\
g_{h\widehat{Z}_A\widehat{Z}_B}=&\pm\varepsilon^2\cos\vartheta\frac{s_W^2m_{Z_A}^2}{v_\text{EW}}M_{Z_AX}M_{Z_BX},\\
g_{h\widehat{X}\widehat{X}}=&\varepsilon^2\frac{s_W^2m_{Z_A}^2}{v_\text{EW}}\cos\vartheta \left( M_{XZ_A}^2+M_{XZ_B}\right).
\end{align}

\section{Useful Formulae}
\label{app:useful}

The $pp\to X$ cross section is obtained by
\begin{equation}
\sigma(pp\to X)=\sum_q\int_{m_X^2/S}^1d\tau L_{\overline{q}q}(\tau)\hat{\sigma}(\overline{q}q\to X)(\hat{s}=\tau S),
\end{equation}
where $S$ is the center of momentum energy of the hadron collider and
\begin{equation}
 L_{\overline{q}q}(\tau)=\int_\tau^1\frac{dx}{x}\left[f_q(x)f_{\overline{q}}\left(\frac{\tau}{x} \right)+f_{\overline{q}}(x)f_q\left(\frac{\tau}{x} \right) \right].
\end{equation}
or
\begin{equation}
L_{\overline{q}q}(\tau)=\int_{-\Upsilon}^{\Upsilon}dy_B\left[f_q(\sqrt{\tau}e^{y_B})f_{\overline{q}}\left(\sqrt{\tau}e^{-y_B} \right)+f_{\overline{q}}(\sqrt{\tau}e^{y_B})f_q\left(\sqrt{\tau}e^{-y_B} \right) \right]\,,
\end{equation}
where $y_B$ is the boost of the dijet ( L.O. partonic) rest frame and $\Upsilon=\min\{y_\text{cut},-\frac{1}{2}\log{\tau}\}$. In either case the $f(x)$ are parton PDFs. We use the MSTW2008 PDFs~\cite{Martin:2009iq} with factorization scale taken to be $m_X$. As our vector are often a little too wide for the narrow width approximation, we use the results from~\cite{Ellis:1991qj} for the partonic cross section
\begin{equation}
\hat{\sigma}(\hat{s})=\frac{\hat{s} m_z^4 \left(g_{q_L}^2+g_{q_R}^2\right)\left(g_{f_L}^2+g_{f_R}^2 \right)}{3 v^4\pi N_c\left[(\hat{s}-m_X^2)^2+\Gamma_X^2m_X^2 \right]}.
\end{equation}

The decay width of a vector $V$ into fermions is,
\begin{equation}
\Gamma(V\to ff)= \frac{N_c  m_V}{24\pi }\sqrt{1-4\frac{m_{f}^2}{m_V^2}}\left[\left({g_L^f}^2+{g_R^f}^2\right)\left(1-\frac{m_{f}^2}{m_V^2}\right)+6g^f_L g^f_R\frac{m_{f}^2}{m_V^2}\right].
\end{equation}
For SM and twin quarks we multiply this by the factor $1+\frac{\alpha_s}{\pi}$ to account for the leading QCD correction. Thus, the production of $\widehat{Z}_B$ will go like $\varepsilon^4$ or $x_A^2\varepsilon^2$.
The decay of a vector $V$ into a pair of $W$s is,
\begin{align}
&\Gamma(V\to WW)= \frac{m_{V}g_{V WW}^2}{192\pi}\left(1-4\frac{m_W^2}{m_V^2}\right)^{\frac{3}{2}}\left(12+20\frac{m_V^2}{m_W^2}+\frac{m_V^4}{m_W^4}\right).
\end{align}
The decay width of a vector $V$ to $\widehat{Z}_{A}h$,
\begin{align}
&\Gamma(V\to \widehat{Z}_A h)= \frac{g_{hV \widehat{Z}_A}^2}{192\pi}\frac{m_{V}}{m_{Z_A}^2}\lambda\left(\frac{m_{Z_A}^2}{m_V^2},\frac{m_h^2}{m_V^2}\right)\left[\lambda^2\left(\frac{m_{Z_A}^2}{m_V^2},\frac{m_h^2}{m_V^2}\right)+12\frac{m_{Z_A}^2}{m_V^2}\right],\\
&\lambda(a,b)=\sqrt{1-2(a+b)+(a-b)^2}.\nonumber
\end{align}
This is modified to other final states in the obvious way.

\bibliography{references}

\end{document}